\documentclass[12pt]{article}

\usepackage{graphicx}

\title{Nonlinear, Band-structure, and Surface Effects in the Interaction of
Charged Particles with Solids}

\author{J. M. Pitarke$^{1,2}$, I. G. Gurtubay$^1$, and V. U. Nazarov$^{3}$\\
$^1$Materia Kondentsatuaren Fisika Saila, UPV/EHU,\\
644 Posta Kutxatila, E-48080 Bilbo, Basque Country\\
$^2$Donostia International Physics Center (DIPC) and\\
Centro Mixto CSIC-UPV/EHU, Basque Country\\
$^3$Department of Physics and Institute for Condensed\\ Matter Theory,
Chonnam National University,\\
Kwangju 500-757, Korea}

\begin{document}
\maketitle

\begin{abstract}
A survey is presented of various aspects of the interaction of charged
particles with solids. In the framework of many-body perturbation theory, we
study the nonlinear interaction of charged particles with a free gas of
interacting electrons; in particular, nonlinear corrections to the stopping
power are analyzed, and special emphasis is made on the separate contributions
that are originated in the excitation of either electron-hole pairs, single
plasmons, or double plasmons. Ab initio calculations of the electronic energy
loss of ions moving in real solids are also presented, and the energy loss of
charged particles interacting with simple metal surfaces is addressed.
\end{abstract}

\newpage

\tableofcontents

\newpage

\section{Introduction}
\label{sec:Intro}

A quantitative description of the interaction of charged particles with solids
is of basic importance in many different theoretical and applied
areas~\cite{echenique1}. When a moving charged particle penetrates a solid
material, it may lose energy to the medium through various elastic and
inelastic collision processes that are based on electron excitation and
nuclear recoil motion in the solid. While energy losses due to nuclear recoil
may become dominant at very low energies of the projectile~\cite{komarov}, in
the case of electrons or ions moving with nonrelativistic velocities that are
comparable to the mean speed of electrons in the solid the most significant
energy losses are due to the generation of electronic excitations, such as
electron-hole (e-h) pairs, collective excitations, i.e., plasmons, and
inner-shell
excitations and ionizations.

In 1913 Niels Bohr published a seminal paper on the energy loss of charged
particles interacting with electrons bound to atoms~\cite{bohr}, which laid
the ground for Bethe's quantum theory of stopping power~\cite{bethe}. The
many-body problem of interacting conduction electrons in metals was
investigated by Bohm and Pines~\cite{bohm} in the so-called {\it jellium}
model, by simply replacing the ionic lattice of the solid by a homogeneous
background which serves to provide neutrality to the system. Within
linear-response theory, the electronic response of conduction electrons to
external charged particles  is determined by a wavevector- and
frequency-dependent longitudinal dielectric function. In the
self-consistent-field, or random-phase, approximation (RPA), the dielectric
function of interacting free electrons was derived by
Lindhard~\cite{lindhard}, and, subsequently, a number of workers have derived
alternative dielectric functions that incorporate various many-body
higher-order local-field corrections~\cite{hubbard,singwi} and band-structure
effects~\cite{adler}. The effect of dissipative processes occurring in a real
metal and conversion of plasmons into multiple e-h pairs has been
allowed phenomenologically by including a damping coefficient in the
dielectric function~\cite{mermin}.

The validity of linear-response theory, which treats the perturbing potential
to lowest order, is not obvious a priori. While lowest-order perturbation
theory leads to energy losses that are proportional to the square of the
projectile charge $Z_1e$, the energy loss of either positive and negative
pions~\cite{barkas} or protons and antiprotons~\cite{andersen1,andersen2} is
known to exhibit a measurable dependence on the sign of the
charge~\cite{ashley1,sung,pitarke1}. Experimentally observed nonlinear
double-plasmon excitations~\cite{double1,double2} are also beyond the realm of
standard linear-response theory~\cite{ashley2,pitarke2,campillo2},
nonlinearities may play an important role in the electronic wake generated by
moving ions in solids~\cite{arnau,pitarke3}, and lowest-order perturbation
theory breaks down when the projectile is capable of carrying bound electrons
with it~\cite{puska}.

The first full nonlinear calculation of the electronic stopping power of an
electron gas was performed by Echenique {\it et al.}~\cite{echenique2}, in the
low-velocity limit. They used a scattering approach to the stopping power, and
the scattering cross sections were calculated for a statically screened
potential which was determined self-consistently by using density-functional
theory (DFT)~\cite{dft1}. These static-screening calculations were then
extended to projectile velocities approaching the Fermi velocity of the
target~\cite{zaremba}. Alternatively, an effective charge can be assigned to
the projectile~\cite{brandt} and nonlinearities can then be investigated
within quadratic-response theory, thereby providing results for arbitrary
projectile velocities~\cite{pitarke4}.

To calculate the electronic stopping power of real solids from the
knowledge of the full band structure of the solid and the corresponding Bloch
eigenfunctions and eigenvalues is a laborious problem. Hence, early theoretical
investigations were based on semiempirical treatments of the electronic
excitations in the solid~\cite{komarov2,desalvo,esbensen,burenkov,crawford}.
Attemps to introduce the full electronic band structure in the evaluation of
the electronic stopping power of alkaline metals for low projectile velocities
include a one-band calculation~\cite{grande} as well as a calculation based on
a linear combination of atomic orbitals~\cite{dorado2}. The low-velocity limit
was also investigated, in the case of silicon, on the basis of a static
treatment of the density response of the solid~\cite{tielens}. Ab initio
band-structure calculations that are based on a full evaluation of the
dynamical density-response of the solid have been carried out only very
recently~\cite{igor0,mathar,igor1}.

In this paper, we summarize recent investigations on the impact of nonlinear,
band-structure, and surface effects in the interaction of charged particles
with solids. We present general procedures to calculate, within many-body
perturbation theory, the nonlinear potential induced by moving ions in an
interacting free-electron gas (FEG), $Z_1^3$ contributions to the stopping
power of a FEG, and double-plasmon excitation probabilities. Self-consistent
calculations of the energy-loss spectra of charged particles moving near a
metal surface are also presented. Finally, we consider the electronic stopping
power of valence electrons in real solids, which we evaluate within random
conditions.

Unless otherwise is stated, atomic units are used throughout this paper, i.e.,
$e^2=\hbar=m_e=1$. The atomic unit of length is the Bohr radius,
$a_0=\hbar^2/m_e^2=0.529\,{\rm \AA}$, the atomic unit of energy is the Hartree,
$1\,{\rm Hartree}=e^2/a_0=27.2\,{\rm eV}$, and the atomic unit of velocity is
the Bohr velocity, $v_0=\alpha\,c=2.19\times 10^8\,{\rm cm}\,{\rm s}^{-1}$,
$\alpha$ and $c$ being the fine structure constant and the velocity of light,
respectively.

\section{Theory}
\label{sec:mytheory}

We consider a recoiless particle of charge $Z_1$~\cite{note1} moving in an
arbitrary inhomogeneous electron system with nonrelativistic velocity ${\bf
v}$, for which retardation effects and radiation losses can be
neglected. The energy that the probe particle loses per unit time due to
electronic excitations in the medium can be written as~\cite{flores}
\begin{equation}\label{flores0}
-{dE\over dt}=-\int d{\bf r}\,\rho^{ext}({\bf r},t)\,{\partial V^{ind}({\bf
r},t)\over\partial t},
\end{equation}
where $\rho^{ext}({\bf r},t)$ represents the probe-particle charge density
\begin{equation}\label{rho}
\rho^{ext}({\bf r},t)=Z_1\,\delta({\bf r}-{\bf r}_0-{\bf v}\,t)
\end{equation}
and
$V^{ind}({\bf r},t)$ is the induced potential. Keeping terms of first and
second order in the external perturbation, time-dependent perturbation theory
yields
\begin{eqnarray}\label{vi}
V^{ind}({\bf r},t)&=&\int d{\bf
r}'\int_{-\infty}^{+\infty} dt'\int_{-\infty}^{+\infty}{d\omega\over
2\pi}\,{\rm
e}^{-i\omega(t-t')}\,\int d{\bf r}_1\int d{\bf
r}_3\,v({\bf r},{\bf r}_1)\,\cr\cr
&\times&\left[\,\chi({\bf r}_1,{\bf r}_3;\omega)
+\int d{\bf
r}''\int_{-\infty}^{+\infty} dt''\int_{-\infty}^{+\infty}
{d\omega'\over 2\pi}{\rm e}^{-i(\omega+\omega')(t'-t'')}\right.\cr\cr
&\times&\left.\int d{\bf r}_2\,Y({\bf r}_1,{\bf r}_2,{\bf
r}_3;\omega,\omega')\,v({\bf r}_2,{\bf r}'')\,\rho^{ext}({\bf r}'',t'')\right]
\,v({\bf r}_3,{\bf r}')\,\rho^{ext}({\bf r}',t'),\nonumber\\
\end{eqnarray}
where $v({\bf r},{\bf r}')=1/|{\bf r}-{\bf r}'|$ is the bare Coulomb
interaction, and $\chi({\bf r}_1,{\bf r}_2;\omega)$ and
$Y({\bf r}_1,{\bf r}_2,{\bf r}_3;\omega,\omega')$ are the so-called linear
and quadratic density-response functions of the electron system:
\begin{equation}\label{chi}
\chi({\bf r}_1,{\bf r}_2;\omega)={1\over\Omega}
\sum_n\left[{\left[\hat\rho({\bf r}_1)\right]_{0n}\left[\hat\rho({\bf
r}_2)\right]_{n0}\over\omega-\omega_{n0}+{\rm i}\eta} -{\left[\hat\rho({\bf
r}_2)\right]_{0n}\left[\hat\rho({\bf r}_1)\right]_{n0}\over
\omega+\omega_{n0}+{\rm i}\eta}\right]
\end{equation}
and
\begin{eqnarray}\label{chi2}
&&Y({\bf r}_1,{\bf r}_2,{\bf
r}_3;\omega,\omega')=-{1\over 2\Omega}\,\sum_{n,l}\left[{\left[\hat\rho({\bf
r}_1)\right]_{0n}\left[\hat\rho({\bf r}_3)\right]_{nl}\left[\hat\rho({\bf
r}_2)\right]_{l0}
\over\left[\omega-\omega_{n0}+{\rm
i}\eta\right]\left[\omega'+\omega_{l0}-{\rm i}\eta\right]}\right.\cr\cr
&&+\left.{\left[\hat\rho({\bf
r}_2)\right]_{0n}\left[\hat\rho({\bf r}_1)\right]_{nl}\left[\hat\rho({\bf
r}_3)\right]_{l0}\over\left[\omega'-\omega_{n0}-{\rm
i}\eta\right]\left[\omega''+\omega_{l0}-{\rm
i}\eta\right]}+{\left[\hat\rho({\bf
r}_3)\right]_{0n}\left[\hat\rho({\bf r}_2)\right]_{nl}\left[\hat\rho({\bf
r}_1)\right]_{l0}\over\left[\omega''-\omega_{n0}-{\rm
i}\eta\right]\left[\omega+\omega_{l0}+{\rm i}\eta\right]}\right.\cr\cr
&&+\left.{\left[\hat\rho({\bf
r}_1)\right]_{0n}\left[\hat\rho({\bf r}_2)\right]_{nl}\left[\hat\rho({\bf
r}_3)\right]_{l0}
\over\left[\omega-\omega_{n0}+{\rm
i}\eta\right]\left[\omega''+\omega_{l0}-{\rm i}\eta\right]}
+{\left[\hat\rho({\bf
r}_3)\right]_{0n}\left[\hat\rho({\bf r}_1)\right]_{nl}\left[\hat\rho({\bf
r}_2)\right]_{l0}\over\left[\omega''-\omega_{n0}-{\rm
i}\eta\right]\left[\omega'+\omega_{l0}-{\rm
i}\eta\right]}\right.\cr\cr
&&+\left.{\left[\hat\rho({\bf
r}_2)\right]_{0n}\left[\hat\rho({\bf r}_3)\right]_{nl}\left[\hat\rho({\bf
r}_1)\right]_{l0}\over\left[\omega'-\omega_{n0}-{\rm
i}\eta\right]\left[\omega+\omega_{l0}+{\rm i}\eta\right]}\right].
\end{eqnarray}
Here, $\Omega$ is the normalization volume, $\hat\rho({\bf r})$ is the
particle-density operator, and $\left[\hat\rho({\bf r})\right]_{nl}$ are matrix
elements taken between the exact many-electron states $\Psi_n$ and $\Psi_l$ of
energy $E_n$ and $E_l$. $\Phi_0$ and $E_0$ represent the exact many-electron
ground state and energy, respectively, $\omega_{nl}=E_n-E_l$,
$\omega''=-(\omega+\omega')$, and $\eta$ is a positive infinitesimal.

The total energy lost by the probe particle is simply
\begin{equation}\label{delta}
-\Delta E=\int_{-\infty}^{+\infty} dt\left(-{dE\over dt}\right),
\end{equation}
which can also be obtained by first considering the probability
$P_{{\bf q},\omega}$ for the probe particle to transfer momentum ${\bf q}$ and
energy $\omega$ to the many-electron system and then multiplying $P_{{\bf
q},\omega}$ by the energy transfer $\omega$ and summing over all ${\bf
q}$ and $\omega$~\cite{pitarke4,teresa1,teresa2,nazarov}.

The many-body ground and excited states of a many-electron system are
{\it unknown}; hence, the exact linear and quadratic density-response functions
are difficult to calculate. In the framework of time-dependent density
functional theory [TDDFT]~\cite{tddft}, the exact density-response functions
are obtained from the knowledge of their noninteracting counterparts and the
exchange-correlation (xc) kernel $f_{xc}({\bf r},{\bf r}',\omega)$ which
equals the second functional derivative of the {\it unknown} xc energy
functional $E_{xc}[n]$. In the so-called time-dependent Hartree approximation
or RPA, the xc kernel is simply taken to be zero.

In the case of a noninteracting Fermi gas, the ground state is obtained by
simply filling all the single-particle states of noninteracting electrons
below the Fermi level. When acting on the ground state, the particle-density
operator $\hat\rho({\bf r})$ produces single-particle transitions in which a
given particle is scattered from some state inside the Fermi sea to a state
outside. Hence, the linear and quadratic density-response functions of
Eqs.~(\ref{chi}) and (\ref{chi2}) reduce to their noninteracting counterparts:
\begin{equation}\label{chi0}
\chi^0({\bf r}_1,{\bf
r}_2;\omega)={2\over\Omega}\sum_{i,j}f_i\left[{\phi_i({\bf r}_1)\phi_j^*({\bf
r}_1)\phi_j({\bf r}_2)\phi_i^*({\bf r}_2)\over \omega-\omega_{ji}+{\rm
i}\eta}-{\phi_i({\bf r}_2)\phi_j^*({\bf r}_2)\phi_j({\bf r}_1)\phi_i^*({\bf
r}_1)\over \omega+\omega_{ji}+{\rm i}\eta}\right]
\end{equation}
and
\begin{eqnarray}\label{chi20}
&&Y^0({\bf r}_1,{\bf r}_2,{\bf
r}_3;\omega,\omega')=-{1\over\Omega}\,\sum_{i,j,k}f_i\left[{\phi_i({\bf
r}_1)\phi_j^*({\bf r}_1)\phi_j({\bf r}_3)\phi_k^*({\bf r}_3)\phi_k({\bf
r}_2)\phi_i^*({\bf r}_2)\over\left[\omega-\omega_{ji}+{\rm
i}\eta\right]\left[\omega'+\omega_{ki}-{\rm i}\eta\right]}\right.\cr\cr
&&+\left.{\phi_i({\bf r}_2)\phi_j^*({\bf r}_2)\phi_j({\bf r}_1)\phi_k^*({\bf
r}_1) \phi_k({\bf r}_3)\phi_i^*({\bf r}_3)\over\left[\omega'-\omega_{ji}-{\rm
i}\eta\right]\left[\omega''+\omega_{ki}-{\rm
i}\eta\right]}
+{\phi_i({\bf r}_3)\phi_j^*({\bf r}_3)\phi_j({\bf
r}_2)\phi_k^*({\bf r}_2)\phi_k({\bf r}_1)\phi_i^*({\bf
r}_1)\over\left[\omega''-\omega_{ji}-{\rm
i}\eta\right]\left[\omega+\omega_{ki}+{\rm i}\eta\right]}\right.\cr\cr
&&+\left.{\phi_i({\bf
r}_1)\phi_j^*({\bf r}_1)\phi_j({\bf r}_2)\phi_k^*({\bf r}_2)\phi_k({\bf
r}_3)\phi_i^*({\bf r}_3)
\over\left[\omega-\omega_{ji}+{\rm
i}\eta\right]\left[\omega''+\omega_{ki}-{\rm i}\eta\right]}
+{\phi_i({\bf
r}_3)\phi_j^*({\bf r}_3)\phi_j({\bf r}_1)\phi_k^*({\bf r}_1)\phi_k({\bf
r}_2)\phi_i^*({\bf r}_2)\over\left[\omega''-\omega_{ji}-{\rm
i}\eta\right]\left[\omega'+\omega_{ji}-{\rm i}\eta\right]}\right.\cr\cr
&&+\left.{\phi_i({\bf
r}_2)\phi_j^*({\bf r}_2)\phi_j({\bf r}_3)\phi_k^*({\bf r}_3)\phi_k({\bf
r}_1)\phi_i^*({\bf r}_1)\over\left[\omega'-\omega_{ji}-{\rm
i}\eta\right]\left[\omega+\omega_{ki}+{\rm i}\eta\right]}\right],
\end{eqnarray}
where $\omega_{ji}=\varepsilon_j-\varepsilon_i$, and $f_i$ are Fermi-Dirac
occupation factors. At zero temperature
$f_i=\Theta(\varepsilon_F-\varepsilon_i)$,  $\varepsilon_F$ being the Fermi
energy, and $\Theta(x)$, the Heaviside step function. The single-particle
states $\phi_i({\bf r})$ and energies $\varepsilon_i$ entering
Eqs.~(\ref{chi0}) and (\ref{chi20}) are usually chosen to be the
eigenfunctions and eigenvalues of an effective Hartree~\cite{fetter},
Kohn-Sham~\cite{gross}, or quasiparticle~\cite{gunnarsson,nekovee}
hamiltonian~\cite{note2}.

In the RPA, the induced potential $V^{ind}({\bf r},t)$ is simply obtained as
the potential induced in a noninteracting Fermi system by both the
probe-particle charge density $\rho^{ext}({\bf r},t)$ and the induced electron
density $\rho^{ind}({\bf r},t)$, i.e.,
\begin{eqnarray}\label{virpa}
V^{ind}({\bf r},t)&=&\int d{\bf
r}'\int_{-\infty}^{+\infty} dt'\int_{-\infty}^{+\infty}{d\omega\over
2\pi}\,{\rm
e}^{-i\omega(t-t')}\,\int d{\bf r}_1\int d{\bf
r}_3\,v({\bf r},{\bf r}_1)\,\cr\cr
&\times&\left\{\,\chi^0({\bf r}_1,{\bf r}_3;\omega)
+\int d{\bf
r}''\int_{-\infty}^{+\infty} dt''\int_{-\infty}^{+\infty}
{d\omega'\over 2\pi}{\rm e}^{-i(\omega+\omega')(t'-t'')}\right.\cr\cr
&\times&\left.\int d{\bf r}_2\,Y^0({\bf r}_1,{\bf r}_2,{\bf
r}_3;\omega,\omega')\,v({\bf r}_2,{\bf r}'')\,\left[\rho^{ext}({\bf r}'',t'')+
\rho^{ind}({\bf r}'',t'')\right]\right\}\cr\cr
&\times&v({\bf r}_3,{\bf r}')\,\left[\rho^{ext}({\bf r}',t')+\rho^{ind}({\bf
r}',t')\right].
\end{eqnarray}
Assuming that
\begin{equation}
V^{ind}({\bf r},t)=\int d{\bf r'} \,v({\bf
r},{\bf r}')\rho^{ind}({\bf r}',t),
\end{equation}
one easily finds that the induced potential $V^{ind}({\bf r},t)$ of
Eq.~(\ref{virpa}) is of the form of Eq.~(\ref{vi}) with the exact interacting
linear and quadratic density-response functions being replaced by the following
integral equations:
\begin{equation}\label{rho2}
\chi({\bf r},{\bf
r}';\omega)=\chi^0({\bf r},{\bf r}';\omega)+\int d{\bf r}_1\int d{\bf
r}_2\chi^0({\bf r},{\bf r}_1;\omega)v({\bf r}_1,{\bf r}_2)\chi({\bf r}_2,{\bf
r}';\omega)
\end{equation}
and
\begin{eqnarray}\label{rho3}
Y({\bf r},{\bf r}',{\bf r}'';\omega,\omega')&=&\int d{\bf r}_1\int d{\bf
r}_2\int d{\bf r}_3K({\bf r},{\bf r}_1;\omega)Y^0({\bf r}_1,{\bf
r}_2,{\bf r}_3;\omega,\omega')\cr\cr
&\times& K({\bf r}_2,{\bf
r}';-\omega')K({\bf r}_3,{\bf r}'';\omega+\omega'),
\end{eqnarray}
where $K({\bf r},{\bf r}';\omega)$ is the so-called inverse dielectric function
\begin{equation}
K({\bf r},{\bf r}';\omega)=\delta({\bf r}-{\bf r}')+\int d{\bf r}_1\,\chi({\bf
r},{\bf r}_1;\omega)\,v({\bf r}_1,{\bf r}').
\end{equation}

\subsection{Uniform electron gas}
\label{sec:basis}
In the case of a uniform FEG, there is translational invariance
in all directions. Hence, Eq.~(\ref{delta}) yields
\begin{equation}\label{uniform}
-\Delta E=L\,\left(-{dE\over dx}\right),
\end{equation}
where $L$ is the normalization length, and $\left(-dE/dx\right)$ is the energy
loss per unit path length of the projectile, i.e., the so-called stopping
power of the electron system:
\begin{eqnarray}\label{eloss}
&&-{dE\over dx}=4\pi\,Z_1^2\int{d{\bf
q}\over(2\pi)^3}\int_0^\infty{d\omega\over 2\pi}\,\omega\,v_{\bf
q}\,\delta\left(\omega-{\bf q}\cdot{\bf v}\right)\cr\cr
&&\times\left[-{\rm Im}K_{{\bf q},\omega}
+2\pi\,Z_1\int{d{\bf
q}_1\over(2\pi)^4}\int_{-\infty}^\infty{d\omega_1\over 2\pi}\,{\rm
Im}Y_{{\bf q},\omega;-{\bf q}_1,-\omega_1}\,v_{{\bf q}_1}\,v_{{\bf q}-{\bf
q}_1}\,\delta\left(\omega_1^0-{\bf q}_1\cdot{\bf v}\right)\right],\nonumber\\
\end{eqnarray}
$v_{\bf q}=4\pi/q^2$, $K_{{\bf q},\omega}$, and $Y_{{\bf q}_1,\omega_1;{\bf
q}_2,\omega_2}$ being Fourier transforms of the bare Coulomb interaction
$v({\bf r},{\bf r}')$, the inverse dielectric function $K({\bf r},{\bf
r}',\omega)$,
and the quadratic density-response function $Y({\bf r},{\bf r}',{\bf
r}'';\omega,\omega')$, respectively.

In the RPA,
\begin{equation}\label{inversep2}
K_{{\bf q},\omega}=1+\chi_{{\bf q},\omega}\,v_{\bf q},
\end{equation}
\begin{equation}\label{chil}
\chi_{{\bf q},\omega}=\chi^0_{{\bf q},\omega}+\chi^0_{{\bf q},\omega}\,
v_{\bf q}\,\chi_{{\bf q},\omega},
\end{equation}
and
\begin{equation}\label{chiq}
Y_{{\bf q}_1,{\bf q}_2;\omega_1,\omega_2}=K_{{\bf q}_1,\omega_1}\,
Y_{{\bf q}_1,{\bf q}_2;\omega_1,\omega_2}^0\,K_{-{\bf
q}_1,-\omega_1}\,K_{{\bf q}_1+{\bf q}_2,\omega_1+\omega_2},
\end{equation}
$\chi^0_{{\bf q},\omega}$ and $Y_{{\bf q}_1,{\bf q}_2;\omega_1,\omega_2}^0$
being Fourier transforms of the noninteracting linear and quadratic
density-response functions $\chi^0({\bf r}_1,{\bf
r}_2;\omega)$ and $Y^0({\bf r}_1,{\bf r}_2,{\bf r}_3;\omega,\omega')$
[see Eqs.~(\ref{chi0}) and (\ref{chi20})]. Noting that for a uniform electron
gas the single-particle states $\phi_i({\bf r})$ entering Eqs.~(\ref{chi0}) and
(\ref{chi20}) are simply plane waves of the form
\begin{equation}
\phi_{\bf k}({\bf r})={1\over\sqrt\Omega}{\rm e}^{i{\bf k}\cdot{\bf r}},
\end{equation}
analytic expressions for both $\chi^0_{{\bf q},\omega}$ and $Y_{{\bf
q}_1,{\bf q}_2;\omega_1,\omega_2}^0$ can be obtained. $\chi^0_{{\bf
q},\omega}$ is the Lindhard function~\cite{lindhard,fetter}. Explicit
expressions for the real and imaginary parts of the noninteracting quadratic
density-response function $Y_{{\bf q}_1,{\bf q}_2;\omega_1,\omega_2}^0$ where
reported in Refs.~\cite{cenni} and~\cite{pitarke1,pitarke4}, respectively, and
an extension to imaginary frequencies was later reported in
Ref.~\cite{ashcroft}.

\subsubsection{High-velocity limit}

For high projectile velocities, the zero-point motion of the electron gas can
be neglected and it can be considered, therefore, as if it were at rest.
Thus, in this approximation all energies $\varepsilon_i$ entering
Eqs.~(\ref{chi0}) and (\ref{chi20}) can be set equal to zero. If one further
assumes
that $v^2>>\omega_p$, $\omega_p$ being the plasmon frequency for which
$K_{0,\omega}$ diverges, Eq.~(\ref{eloss}) is found to yield~\cite{pitarke4}
\begin{equation}\label{limit0}
-{dE\over dx}\approx{4\pi nZ_1^2\over v^2}(L_0+Z_1L_1),
\end{equation}
where
\begin{equation}
L_0=\ln{2v^2\over\omega_p}
\end{equation}
and
\begin{equation}
L_1=1.42{\pi\omega_p\over v^3}\,\ln{2v^2\over 2.13\omega_p},
\end{equation}
$n$ being the density of free electrons in the target. The first term, which
for a relatively low projectile charge $Z_1$ gives the main contribution to
the stopping power, is proportional to $Z_1^2$ and has the same form as the
Bethe formula for the inelastic stopping power of atoms~\cite{bethe} as long
as the plasma frequency $\omega_p$ is replaced by the mean excitation energy
of electrons in the atom. The second term, which originates the so-called
Barkas effect, i.e., the dependence of the stopping power on the sign of the
projectile charge, has been found to yield excellent agreement with
stopping-power measurements at high velocities~\cite{pitarke1,grande2}.

\subsubsection{Double-plasmon excitation}

In the RPA, the linear-response contribution to the stopping power of
Eq.~(\ref{eloss}), which is proportional to $Z_1^2$, is fully originated in the
creation of single e-h pairs and plasmons. Furthermore, the
contribution to the actual (beyond RPA) $Z_1^2$ stopping power that is due to
coherent multiple excitations such as double plasmons is expected to be
relatively small. Nevertheless, accurate measurements of electron energy-loss
spectra showed evidence for the existence of coherent double-plasmon
excitations~\cite{double1,double2}.

In the framework of many-body perturbation theory, one first defines the
scattering matrix $S$ as a time-ordered exponential in terms of the
perturbing hamiltonian and field operators~\cite{fetter}. Then, one considers
the matrix elements corresponding to the process in which the recoiless probe
particle carries the system either from an initial state $a_i^+\Phi_0$ to a
final state $a_f\Phi_0$ (single excitation) or from an initial state
$a_{i_1}^+a_{i_2}^+\Phi_0$ to a final state $a_{f_1}^+a_{f_2}^+\Phi_0$ (double
excitation):
\begin{equation}\label{14p}
S_{f,i}={<\Phi_0|a_fSa_i^+|\Phi_0>\over
<\Phi_0|S|\Phi_0>}
\end{equation}
and
\begin{equation}\label{15p}
S_{f_1,f_2;i_1,i_2}={<\Phi_0|a_{f_1}a_{f_2}Sa_{i_1}^+a_{i_2}^+|\Phi_0>\over
<\Phi_0|S|\Phi_0>},
\end{equation}
where $a_i$ and $a_i^+$ are annihilation and creation operators,
respectively, and $\Phi_0$ represents the vacuum state. Finally, one
calculates the probability for the probe particle to transfer momentum ${\bf
q}$ and energy $\omega$ to the many-electron system by moving either one, two,
or more particles from inside the Fermi sea to outside:
\begin{eqnarray}\label{probability}
P_{{\bf q},\omega}&=&2\sum_{\bf s}f_{\bf s}\sum_{\bf
p} (1-f_{\bf p})\left|S_{f,i}\right|^2\delta^4_{q,p-s} +4\sum_{{\bf
q}_1}\sum_{\omega_1}\sum_{{\bf s}_1}f_{{\bf s}_1} \sum_{{\bf s}_2}f_{{\bf
s}_2} \sum_{{\bf p}_1}(1-f_{{\bf p}_1})\cr\cr
&\times&\sum_{{\bf p}_2}
(1-f_{{\bf p}_2})
\left|S_{f_1,f_2;i_1,i_2}\right|^2\delta^4_{q_1,p_1-s_1}
\delta^4_{q-q_1,p_2-s_2}+\cdots,
\end{eqnarray}
$\delta^4_{q,q'}=\delta^3_{{\bf q},{\bf q}'}\delta_{\omega,\omega'}$ being
the symmetric Kronecker $\delta$ symbol. If the probe-particle is not a
heavy particle, energy conservation should be ensured by introducing recoil
into the argument of the delta functions. If the probe is an electron, a step
function should also be introduced to ensure that the probe electron does not
lose enough energy to fall below the Fermi level.

The stopping power of the electron system is obtained by first multiplying
the probability $P_{{\bf q},\omega}$ of Eq.~(\ref{probability}) by the energy
transfer $\omega$ and then summing over all ${\bf q}$ and $\omega$:
\begin{equation}\label{delta2}
-{dE\over dx}={1\over L}\sum_{\bf q}\sum_\omega\,\omega\,P_{{\bf q},\omega}.
\end{equation}
A careful analysis of the various contributions to the probability $P_{{\bf
q},\omega}$ that are proportional to $Z_1^2$ and $Z_1^3$ yields, to third
order in the RPA screened interaction $v_{\bf q}\,K_{{\bf q},\omega}$, an
expression for the stopping power that exactly coincides with
Eq.~(\ref{eloss})~\cite{teresa2}.  Alternatively, the inverse mean free path
is obtained as follows
\begin{equation}\label{lambda}
\lambda^{-1}={1\over
L}\sum_{\bf q}\sum_\omega\,P_{{\bf q},\omega}.
\end{equation}

The lowest-order $Z_1^2$ contribution to the probability $P_{{\bf
q},\omega}^{2p}$ for a probe-electron ($Z_1=-1$) to excite a double plasmon,
which is of fourth order in the RPA screened interaction $v_{\bf q}K_{{\bf
q},\omega}$, is found to be given by the following expression:
\begin{eqnarray}\label{p2p}
P_{{\bf q},\omega}^{2p}&=&16\pi Z_1^2\,v_{\bf q}^{-2}K_{{\bf q},\omega}^{-2}
\sum_{{\bf q}_1}v_{{\bf q}_1}v_{{\bf q}-{\bf
q}_1}\int_0^{\omega}{d\omega_1\over 2\pi}\,{\rm Im}K_{{\bf
q}_1,\omega_1}\,{\rm Im}K_{{\bf q}-{\bf q}_1,\omega-\omega_1}\cr\cr
&\times& |Y_{{\bf q},{\bf q}_1;\omega,\omega_1}|^2 \,\delta(q^0-p^0+
\omega_{{\bf v}-{\bf q}})\,\Theta(\omega_{{\bf v}-{\bf q}}-\varepsilon_F),
\end{eqnarray}
where $\omega_{\bf k}=k^2/2$. If one approximates both the linear and
quadratic density-response functions entering Eq.~(\ref{p2p}) by their
low-${\bf q}$ limits, one only keeps the high-velocity limit of this
probability in an expansion in terms of the inverse velocity, and one
introduces this limit into Eq.~(\ref{lambda}), one obtains the following
result for the $Z_1^2$ contribution to the inverse mean free path that is due
to the excitation of a double plasmon~\cite{pitarke2,note3}:
\begin{equation}\label{limit}
\lambda_{2p}^{-1}\approx 0.164\,Z_1^2\,{\sqrt{r_s}\over
36\pi v^2},
\end{equation}
$r_s$ being the so-called electron-density
parameter $r_s$ defined by the relation $1/n=(4\pi/3)(r_s\,a_0)^3$. Numerical
study shows~\cite{campillo2} that introduction of the full RPA linear and
quadratic density-response functions into Eq.~(\ref{p2p}) yields a result for
$\lambda_{2p}^{-1}$ which has, in the high-velocity limit, the same velocity
dependence as the approximation of Eq.~(\ref{limit}), though for $r_s=2.07$
the full RPA $\lambda_{2p}^{-1}$ is found to be larger than this approximation
by a factor of $2.16$.

\subsection{Bounded electron gas}

In the case of charged particles moving inside a solid, nonlinear effects are
known to be crucial in the interpretation of energy-loss measurements.
However, these corrections have been shown to be less important when the
charged particle moves outside the solid~\cite{bergara}. Hence, in the case
of a bounded three-dimensional electron gas we restrict the calculations to
linear-response theory. Assuming translational invariance in two directions,
which we take to be normal to the $z$ axis, to first order in the external
perturbation (linear-response theory) the energy loss of Eq.~(\ref{delta}) may
be expressed as follows~\cite{aran}
\begin{eqnarray}\label{general}
-\Delta E&=&-{Z_1^2\over\pi}\int {d{\bf
q}_\parallel\over(2\pi)^2}\int_{-\infty}^{+\infty} dt\int_{-\infty}^{+\infty}
dt' \int_0^\infty d\omega\,\omega\cr\cr &\times&{\rm e}^{-i(\omega-{\bf
q}_\parallel\cdot{\bf v}_\parallel)(t-t')}\,{\rm Im}W[z(t),z(t');{\bf
q}_\parallel,\omega],
\end{eqnarray}
where ${\bf q}_\parallel$ and ${\bf v}_\parallel$ are components of the
momentum transfer and the velocity in the plane of the surface, $z(t)$
represents the position of the projectile relative to the surface, and
\begin{equation}\label{screened2}
W(z,z';{\bf q}_\parallel,\omega)=v(z,z',{\bf
q}_\parallel) +\int dz_1\int dz_2\,v(z,z_1;{\bf q}_\parallel)
\,\chi(z_1,z_2;{\bf q}_\parallel,\omega)\,v(z_2,z';{\bf q}_\parallel),
\end{equation} $v(z,z';{\bf q}_\parallel)=(2\pi/q_\parallel){\rm
e}^{-q_\parallel\,|z-z'|}$ and $\chi(z,z';{\bf q}_\parallel,\omega)$ being
two-dimensional Fourier transforms of the bare Coulomb interaction and the
density-response function, respectively.

In the RPA,
\begin{eqnarray}\label{chiz}
\chi(z,z';{\bf q}_\parallel,\omega)&=&
\chi^0(z,z';{\bf q}_\parallel,\omega)+\int dz_1\int{\rm
d}z_2\chi^0(z,z_1;{\bf q}_\parallel,\omega)\cr\cr
&\times&v(z_1,z_2;{\bf q}_\parallel)\,\chi(z_2,z';{\bf q}_\parallel,\omega),
\end{eqnarray}
where
\begin{equation}\label{chiz0}
\chi^0(z,z';{\bf q}_\parallel,\omega)=2\,\sum_{i,j}\phi_i(z)\phi_{j}^*(z)
\phi_{j}(z')\phi_i^*(z')
\,\int{{\rm d}{\bf
k}_\parallel\over(2\pi)^2}\,{f_i-f_j\over
E_i-E_{j}+(\omega+i\eta)}.
\end{equation}
Here,
\begin{equation}\label{4}
E_i=\varepsilon_i+{{\bf k}_\parallel^2\over 2}
\end{equation}
and
\begin{equation}\label{4p}
E_{j}=\varepsilon_{j}+{({\bf k}_\parallel+{\bf q}_\parallel)^2\over 2},
\end{equation}
the wave functions $\phi_i(z)$ and energies $\varepsilon_i$, which describe
motion normal to the surface, being the eigenfunctions and eigenvalues of
a one-dimensional Hartree, Kohn-Sham, or quasiparticle hamiltonian.

Eq.~(\ref{general}) gives the energy that a charged particle moving with
constant velocity along an arbitrary trajectory loses due to electronic
excitations in an electron system that is translationally invariant in two
directions, as occurs in the case of a simple metal surface modeled by jellium.

\subsubsection{Parallel trajectory}

In the glancing incidence geometry ions penetrate into the material, they skim
the outermost layer of the solid, and are then repelled by a repulsive,
screened Coulomb potential, as discussed by Gemmell~\cite{Gemmell}. Through use
of the appropriate effective potentials the ion trajectory $z(t)$ can be
calculated and the energy loss is then obtained from Eq.~(\ref{general}).
Under extreme grazing-incidence conditions, incident charged particles can be
assumed to move with constant velocity ${\bf v}$ along a definite trajectory
at a fixed distance $z$ from a jellium surface. Eq.~(\ref{general}) then
yields Eq.~(\ref{uniform}), as in the case of charged particles moving in a
uniform electron gas, but with the energy loss per unit path length now being
given by the following expression~\cite{aran}:
\begin{equation}\label{stop}
-{dE\over dx}=-{2\over v}\,Z_1^2\int{d{\bf
q}_\parallel\over(2\pi)^2}\int_0^\infty d\omega\,\omega\,{\rm
Im}W(z,z;{\bf q}_\parallel,\omega)\,\delta(\omega-{\bf q}_\parallel\cdot{\bf
v}).
\end{equation}

\subsubsection{High-velocity limit}

At high velocities, the energy-loss spectrum for charged particles moving
outside a solid is known to be dominated by long-wavelength ($q_\parallel\to
0$) surface-plasmon excitations~\cite{ritchie}. In this long-wavelength limit,
the imaginary part of Eq.~(\ref{screened2}) yields~\cite{liebsch}
\begin{equation}\label{winf}
{\rm Im}\,W(z,z';q_\parallel,\omega)=-{\pi^2\over
q_\parallel}\,e^{-q_\parallel(z+z')}\,\omega_s\,\delta(\omega-\omega_s),
\end{equation}
where $\omega_s=\omega_p/\sqrt{2}$, $\omega_p={(4\pi n)}^{1/2}$ being the
classical plasma frequency of a uniform electron gas of density $n$.

Introducing Eq.~(\ref{winf}) into Eq.~(\ref{stop}), one easily reproduces the
classical expression of Echenique and Pendry~\cite{pendry}:
\begin{equation}\label{stopcl}
-{dE\over dx}=Z_1^2\,{\omega_s^2\over v^2}\, {\rm K}_0(2\,\omega_s\,z/v),
\end{equation}
where ${\rm K}_0$ is the zero-order modified Bessel function. For large values
of $z$ ($z>>v/\omega_s$), Eq.~(\ref{stopcl}) reduces to
\begin{equation}\label{stopcllargez}
-{dE\over dx}=Z_1^2\,{\omega_s\over
2\,v}\, \sqrt{\pi\,\omega_s/z\,v}\,\,e^{-2\,\omega_s\,z/v}.
\end{equation}

\subsection{Periodic crystals}

For a peridic crystal, we introduce the following Fourier expansion of the
linear density-response function:
\begin{equation}\label{eq8}
\chi({\bf r},{\bf r}';\omega)={1\over\Omega}\,\sum_{\bf q}^{BZ}\sum_{{\bf
G},{\bf G}'}{\rm e}^{{\rm i}({\bf q}+{\bf G})\cdot{\bf r}}{\rm
e}^{-{\rm i}({\bf q}+{\bf G}')\cdot{\bf r}'}\chi_{{\bf G},{\bf G}'}({\bf
q},\omega),
\end{equation}
where the first sum runs over ${\bf q}$ vectors within the first Brillouin
zone (BZ), and ${\bf G}$ and ${\bf G}'$ are reciprocal lattice vectors.

In the RPA,
\begin{equation}\label{eq28}
\chi_{{\bf G},{\bf G}'}({\bf q},\omega)=\chi_{{\bf G},{\bf G}'}^0({\bf
q},\omega)+ \sum_{{\bf G}''}\chi_{{\bf G},{\bf G}''}^0({\bf q},\omega)v_{{\bf
q}+{\bf G}''}\,\chi_{{\bf G}'',{\bf G}'}({\bf q},\omega),
\end{equation}
where $v_{{\bf q}+{\bf G}}$ and $\chi^0_{{\bf G},{\bf
G}'}({\bf q},\omega)$ represent the Fourier coefficients of the
bare Coulomb interaction $v({\bf r},{\bf r}')$ and the noninteracting
density-response function $\chi^0({\bf r},{\bf r}';\omega)$, respectively:
\begin{equation}
v_{{\bf q}+{\bf G}}={4\pi\over|{\bf q}+{\bf G}|^2}
\end{equation}
and
\begin{eqnarray}\label{eq9}
\chi_{{\bf G},{\bf G}'}^0({\bf
q},\omega)&=&{1\over\Omega}\sum_{\bf k}^{BZ}\sum_{n,n'}\,(f_{{\bf k},n}-f_{{\bf
k}+{\bf q},n'})\cr\cr
&\times&{\langle\phi_{{\bf k},n}|e^{-{\rm i}({\bf q}+{\bf
G})\cdot{\bf r}}|\phi_{{\bf k}+{\bf q},n'}\rangle \langle\phi_{{\bf k}+{\bf
q},n'}|e^{{\rm i}({\bf q}+{\bf G}')\cdot{\bf r}}|\phi_{{\bf k},n}\rangle
\over E_{{\bf k},n}-E_{{\bf k}+{\bf q},n'} +\hbar(\omega + {\rm
i}\eta)},
\end{eqnarray}
$\phi_{{\bf k},n}$ and $E_{{\bf k},n}$ being Bloch
eigenfunctions and eigenvalues of a three-dimensional Hartree, Kohn-Sham, or
quasiparticle hamiltonian.

The stopping power of a periodic crystal is obtained by first introducing
Eqs.~(\ref{rho}) and (\ref{vi}) into Eq.~(\ref{flores0}), and then introducing
Eq.~(\ref{flores0}) into Eq.~(\ref{delta}). Within linear-response theory,
i.e., to first order in the external perturbation, the result is of the form
of  Eq.~(\ref{uniform}) with the energy loss per unit path length being given
by the following expression:
\begin{equation}\label{position}
\left[-{dE\over dx}\right]_{\bf b}=-{2\,Z_1^2\over v\,\Omega}\,\sum_{\bf
q}^{\rm BZ}\sum_{\bf G}\sum_{{\bf G}_\perp'}\,\omega\,{\rm e}^{{\rm
i}{\bf G}_\perp'\cdot{\bf b}}\,
v_{{\bf q}+{\bf G}+{\bf G}_\perp'}\,{\rm Im}\,K_{{\bf
G},{\bf G}+{\bf G}_\perp'}({\bf q},\omega),
\end{equation}
where ${\bf b}$ is the impact vector of the
projectile, the sum $\sum_{{\bf G}_\perp'}$ is restricted to those
reciprocal-lattice vectors that are perpendicular to the projectile velocity
(${\bf G}_\perp'\cdot{\bf v}=0$), $\omega=({\bf q}+{\bf G})\cdot{\bf v}$, and
$K_{{\bf G},{\bf G}'}({\bf q},\omega)$ represent the Fourier coefficients of
the inverse dielectric function $K({\bf r},{\bf r}';\omega)$:
\begin{equation}\label{eq27}
K_{{\bf G},{\bf G}'}({\bf q},\omega)=\delta_{{\bf
G},{\bf G}'}+v_{{\bf q}+{\bf G}}\,\chi_{{\bf G},{\bf G}'}({\bf q},\omega).
\end{equation}

The stopping power of Eq.~(\ref{position}) is the so-called position-dependent
stopping power. The most important contribution to this quantity is provided by
the term ${\bf G}_\perp'=0$, the magnitude of the other terms depending on the
direction of the velocity. For a few highly symmetric or {\it channeling}
directions non-negligible corrections to the ${\bf G}_\perp'=0$ contribution
are found, thus exhibiting the characteristic anisotropy of the
position-dependent stopping power. However, for those directions for which the
condition of ${\bf G}_\perp'\cdot{\bf v}=0$ is never satisfied only the ${\bf
G}_\perp'=0$ term contributes and one finds
\begin{equation}\label{random}
\left[-{dE\over dx}\right]_{\rm random}=-{2\,Z_1^2\over v\,\Omega}\,\sum_{\bf
q}^{\rm BZ}\sum_{\bf G}\,\omega\,v_{{\bf q}+{\bf G}}\,{\rm Im}\,K_{{\bf G},{\bf
G}}({\bf q},\omega).
\end{equation}
This is the so-called random stopping power, which is also obtained as the
average over impact parameters of the position-dependent stopping power of
Eq.~(\ref{position}). For simple metals like Al the diagonal elements of the
inverse dielectric matrix $K_{{\bf G},{\bf G}}({\bf q},\omega)$ are rather
isotropic, in which case there is little dependence of the random stopping
power on the direction of the projectile velocity.

\section{Results}
\label{results}

In this section, we review existing calculations of the stopping power of both
an infinite (uniform) and a semi-infinite (bounded) free gas of interacting
electrons, double-plasmon inverse mean free paths, and the linear
(lowest-order) first-principles stopping power of real Al and Si. New
calculations of the various contributions to the stopping power due to
the excitation of e-h pairs and plasmons are also reported, existing
double-plasmon calculations are extended to low velocities where both recoil
and the probe-particle statistics play a role, and existing first-principles
calculations of the stopping power of Al and Si are extended by allowing
transitions with large values of the momentum transfer.

\subsection{Uniform electron gas}

Here we consider a uniform gas of interacting electrons, the electron density
$n$ being equal to the average electron density of valence electrons in
aluminum metal ($r_s=2.07$), for which the Fermi momentum [$q_F=(3\pi^2
n)^{1/3}$] and bulk plasma frequency [$\omega_p=(4\pi n)^{1/2}$] are
$q_{F}=0.927\,a_0^{-1}$ and $\omega_p=15.8$ eV, respectively. We set $Z_1=1$
(probe bare protons) for the stopping-power calculations and $Z_1=-1$ (probe
electrons) for the calculations of the double-plasmon inverse mean free path.
Our results can then be easily extended to arbitrary values of $Z_{1}$, as
linear and quadratic contributions to the stopping power and the inverse mean
free path are proportional to $Z_1^2$ and $Z_1^3$, respectively.

\subsubsection{Stopping power}

\begin{figure}
\begin{center}
\includegraphics[width=0.65\textwidth]{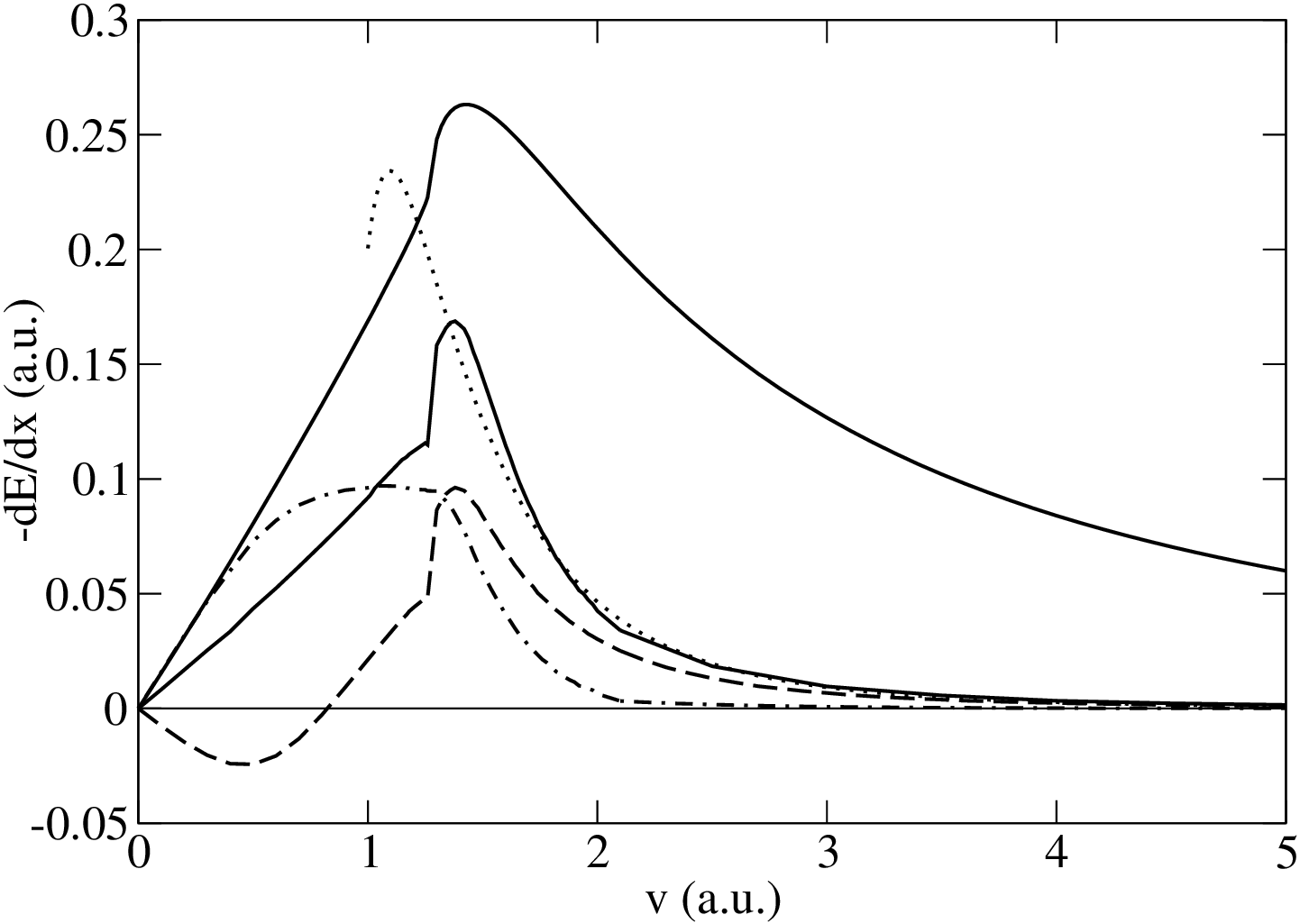}
\caption{\label{fig1}
Solid lines represent RPA $Z_1^2$ and $Z_1^3$ contributions to the stopping
power of Eq.~(\ref{eloss}), for $Z_1=1$ and $r_s=2.07$, as a function of the
velocity of the projectile. Dashed and dashed-dotted lines represent the RPA
$\left[-dE/dx\right]^1$ and $\left[-dE/dx\right]^2$ contributions
to the $Z_1^3$ stopping power, respectively. The dotted line is the
high-velocity
limit dictated by the second term of Eq.~(\ref{limit0}).}
\end {center}
\end{figure}

The second-order stopping power of a homogeneous FEG is given by
Eq.~(\ref{eloss}).  In Fig.~\ref{fig1} we show, as a function of the
projectile velocity, our full RPA calculations for the separate $Z_1^2$ and
$Z_1^3$ contributions to the stopping  power of Eq.~(\ref{eloss}). We note
that both $Z_1^2$ and $Z_1^3$ contributions to the stopping power exhibit a
linear dependence on the projectile velocity up to velocities approaching the
stopping maximum. This linear dependence is also exhibited by full nonlinear
DFT calculations of the stopping power of a FEG~\cite{echenique2} and
by recent measurements of the electronic energy loss of protons and
antiprotons~\cite{andersen2}.

The linear dependence of the $Z_1^3$ contribution to the stopping power is a
consequence of two competing effects. On the one hand, there is the effect of
one-step single excitations, like those entering the $Z_1^2$ contribution to
the stopping power, but now generated by the quadratically screened potential
of the probe particle. On the other hand, there is the effect of second-order
two-step single excitations generated by the linearly screened probe potential.
In the RPA, these contributions to the $Z_1^3$ effect are given by the
following expressions~\cite{pitarke4}:
\begin{eqnarray}\label{eloss1}
\left[-{dE\over dx}\right]^1&=&-{4\over v}\,Z_1^3\int{d{\bf
q}\over(2\pi)^3}\int_0^\infty{d\omega\over 2\pi}\,\omega\,v_{\bf
q}\,{\rm Im}\,K_{{\bf q},\omega}\,\delta\left(\omega-{\bf q}\cdot{\bf v}\right)
\int{d{\bf q}_1\over(2\pi)^4}\int_{-\infty}^\infty{d\omega_1\over
2\pi}\cr\cr
&\times&{\rm Re}\,Y_{{\bf q},\omega;-{\bf q}_1,-\omega_1}^{TO}\,v_{{\bf
q}_1}\,{\rm Re}\,K_{{\bf q}_1,\omega_1}\,v_{{\bf q}-{\bf q}_1}\,{\rm
Re}\,K_{{\bf q}-{\bf q}_1,\omega-\omega_1}\delta\left(\omega_1^0-{\bf
q}_1\cdot{\bf v}\right)\nonumber\\
 \end{eqnarray}
and
\begin{eqnarray}\label{eloss2}
\left[-{dE\over dx}\right]^2&=&-{4\over v}\,Z_1^3\int{d{\bf
q}\over(2\pi)^3}\int_0^\infty{d\omega\over 2\pi}\,\omega\,v_{\bf
q}\,{\rm Re}\,K_{{\bf q},\omega}\,\delta\left(\omega-{\bf q}\cdot{\bf v}\right)
\int{d{\bf q}_1\over(2\pi)^4}\int_{-\infty}^\infty{d\omega_1\over
2\pi}\cr\cr
&\times&H_{{\bf q},\omega;-{\bf q}_1,-\omega_1}\,v_{{\bf
q}_1}\,{\rm Re}\,K_{{\bf q}_1,\omega_1}\,v_{{\bf q}-{\bf q}_1}\,{\rm
Re}\,K_{{\bf q}-{\bf q}_1,\omega-\omega_1}\delta\left(\omega_1^0-{\bf
q}_1\cdot{\bf v}\right),\nonumber\\
 \end{eqnarray}
where $Y_{{\bf q},\omega;{\bf q}_1,\omega_1}^{TO}$ is the time-ordered
counterpart of the retarded non-interacting quadratic density-response function
$Y_{{\bf q},\omega;{\bf q}_1,\omega_1}^0$, and
\begin{equation}\label{h}
H_{{\bf q},\omega;{\bf q}_1,\omega_1}={2\pi{\rm sgn}(\omega) P}\int{{\rm
d}^3{\bf k}\over (2\pi)^3}f_{\bf k}\left[{\delta(\omega+\omega_{\bf
k}-\omega_{{\bf k}+{\bf q}})\over \omega_1+\omega_{\bf k}-\omega_{{\bf k}+{\bf
q}_1}}+{\delta(\omega^0-\omega_{\bf k}+\omega_{{\bf k}+{\bf q}}) \over
\omega_2-\omega_{\bf k}+\omega_{{\bf k}+{\bf q}_2}}\right],\nonumber\\
\end{equation}
with ${\bf q}_2={\bf q}-{\bf q}_1$ and $\omega_2=\omega-\omega_1$.
Double excitations also contribute, within RPA, to the $Z_1^3$ stopping power
of Eq.~(\ref{eloss}), but they are found to be zero in the low and high
velocity limits and small at intermediate velocities.

The RPA $\left[-dE/dx\right]^1$ and $\left[-dE/dx\right]^2$ contributions to
the stopping power of a FEG are represented in Fig.~\ref{fig1} by dashed and
dashed-dotted lines, respectively. The $\left[-dE/dx\right]^2$ contribution
from losses to two-step single excitations (dashed-dotte line) is very small at
high projectile velocities where the velocity distribution of target electrons
can be neglected. At high velocities, the surviving $\left[-dE/dx\right]^1$
contribution approaches the total $Z_1^3$ stopping power, which is very well
reproduced by the second term of Eq.~(\ref{limit0}) (dotted line of
Fig.~\ref{fig1}).

\begin{figure}
\begin{center}
\includegraphics[width=0.65\textwidth]{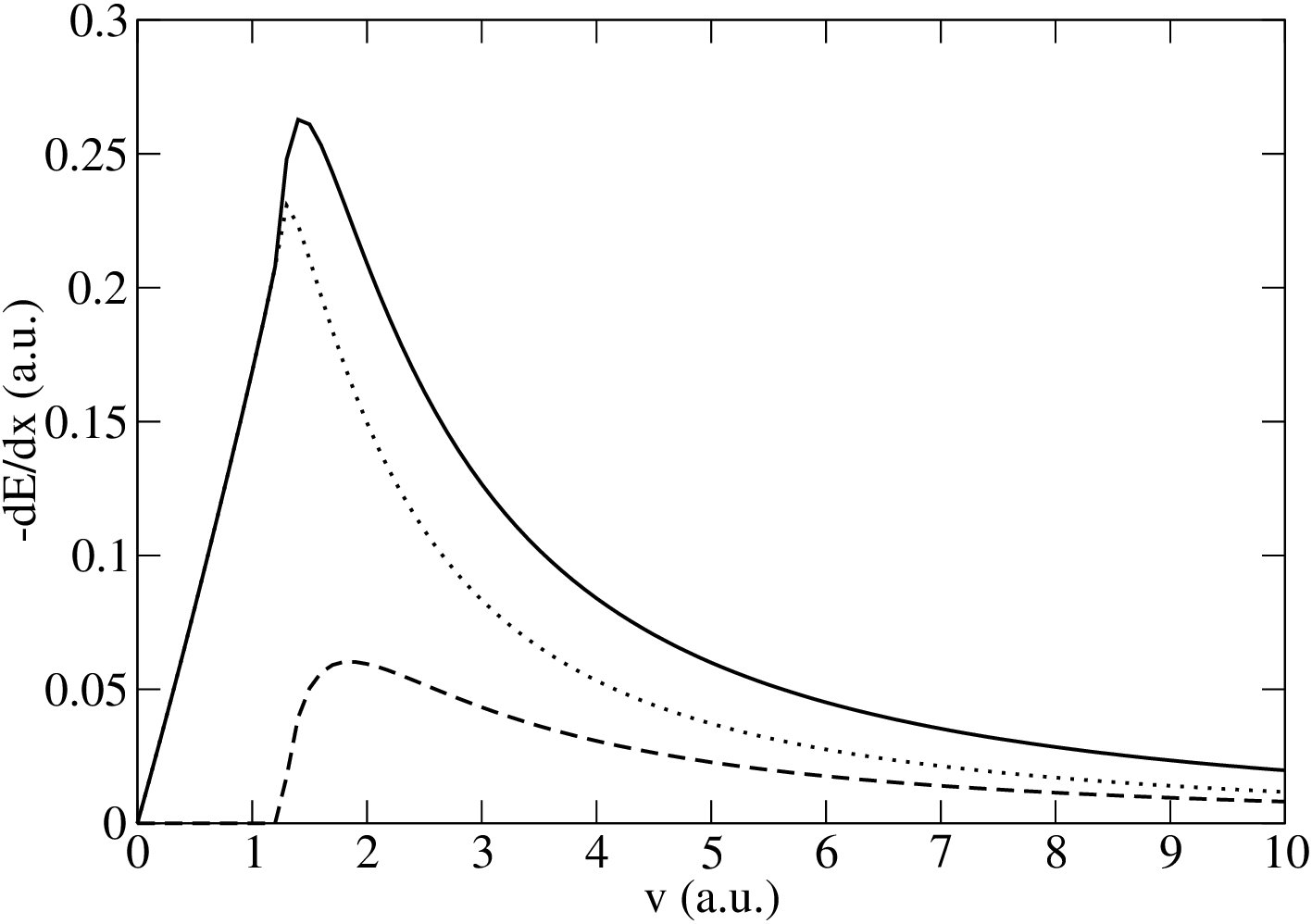}
\caption{\label{fig2}
The solid line represents the RPA $Z_1^2$ (linear) contribution to the stopping
power of Eq.~(\ref{eloss}), for $Z_1=1$ and $r_s=2.07$, as a function of the
velocity of the probe particle. Dashed and dotted lines represent
contributions from the excitation of plasmons and e-h pairs, respectively.}
\end{center}
\end{figure}

Now we focus on the role that the excitation of e-h pairs and plasmons plays in
the energy loss process. First, we restrict our calculations to the $Z_1^2$
(linear) contribution to the stopping power of a FEG. Fig.~\ref{fig2}
exhibits the separate RPA e-h pair and plasmon contributions to the $Z_1^2$
stopping
power of Eq.~(\ref{eloss}). This figure shows that the contribution from losses
to
plasmons is smaller for all projectile velocities than the contribution from
losses to e-h pairs, which is especially true at high electron densities,
although both contributions coincide in the high-velocity limit. This is the
equipartition rule, which appears straightforwardly in the static-electron gas
approximation. In this approximation, plasmon and e-h pair contributions to
the energy loss are typically separated according to whether the momentum
transfer is below ($q<q_c$) or above ($q>q_c$) the critical momentum
$q_c$ where the plasmon dispersion enters the e-h pair excitation spectrum.
For an electron gas not at rest, the equipartition rule was formulated by
Lindhard and Winther~\cite{winther}.

\begin{figure}
\begin{center}
\includegraphics[width=0.65\textwidth]{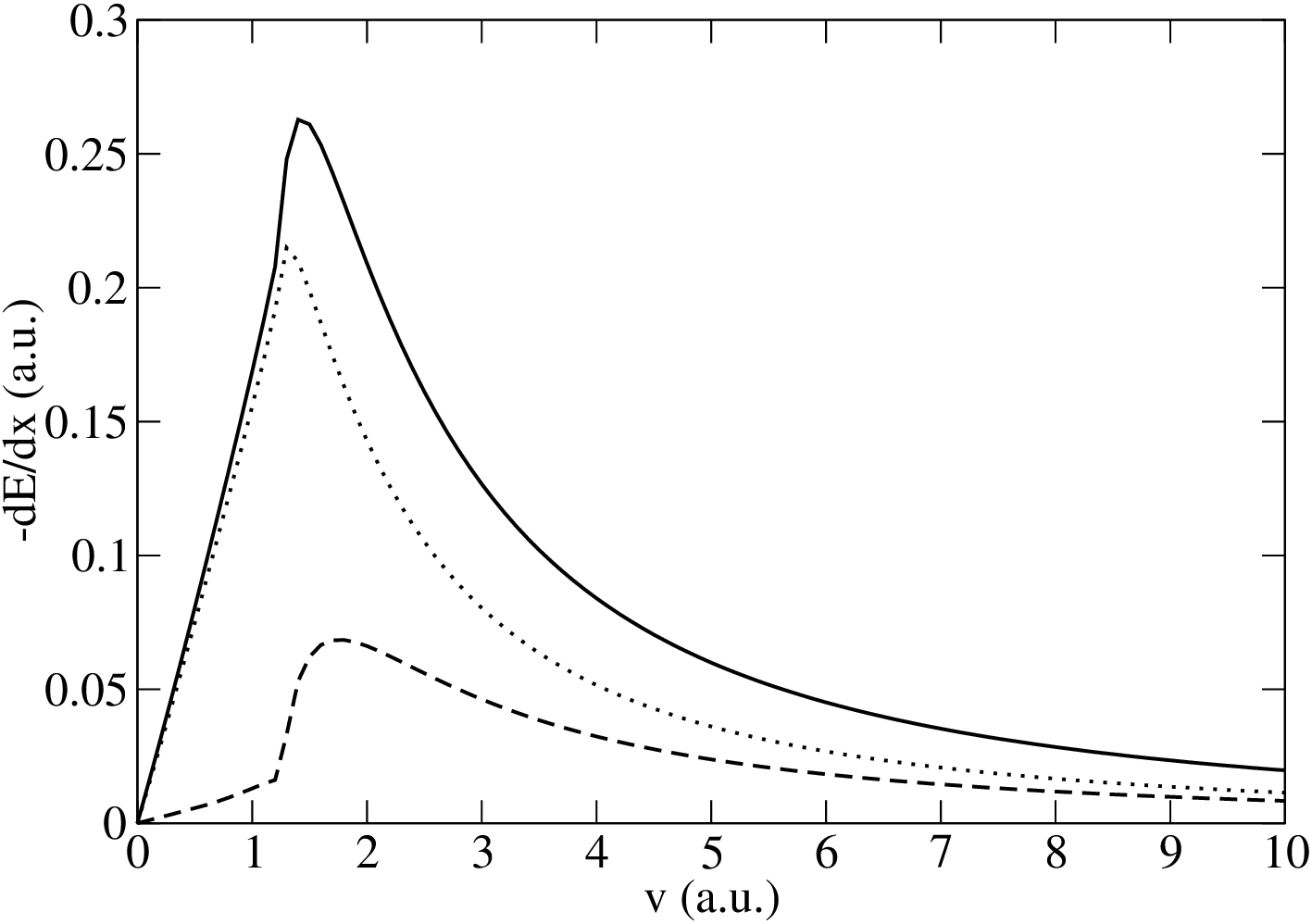}
\caption{\label{fig3}
The solid line represents the RPA $Z_1^2$ (linear) contribution to the
stopping power of Eq.~(\ref{eloss}), for $Z_1=1$ and $r_s=2.07$, as a function
of the velocity of the probe particle. Dashed and dotted lines represent
contributions from momentum transfers below ($q<q_c$) and above ($q>q_c$) the
critical momentum $q_c$ where the plasmon dispersion enters the e-h pair
continuum.}
\end{center}
\end{figure}

\begin{figure}
\begin{center}
\includegraphics[width=0.65\textwidth]{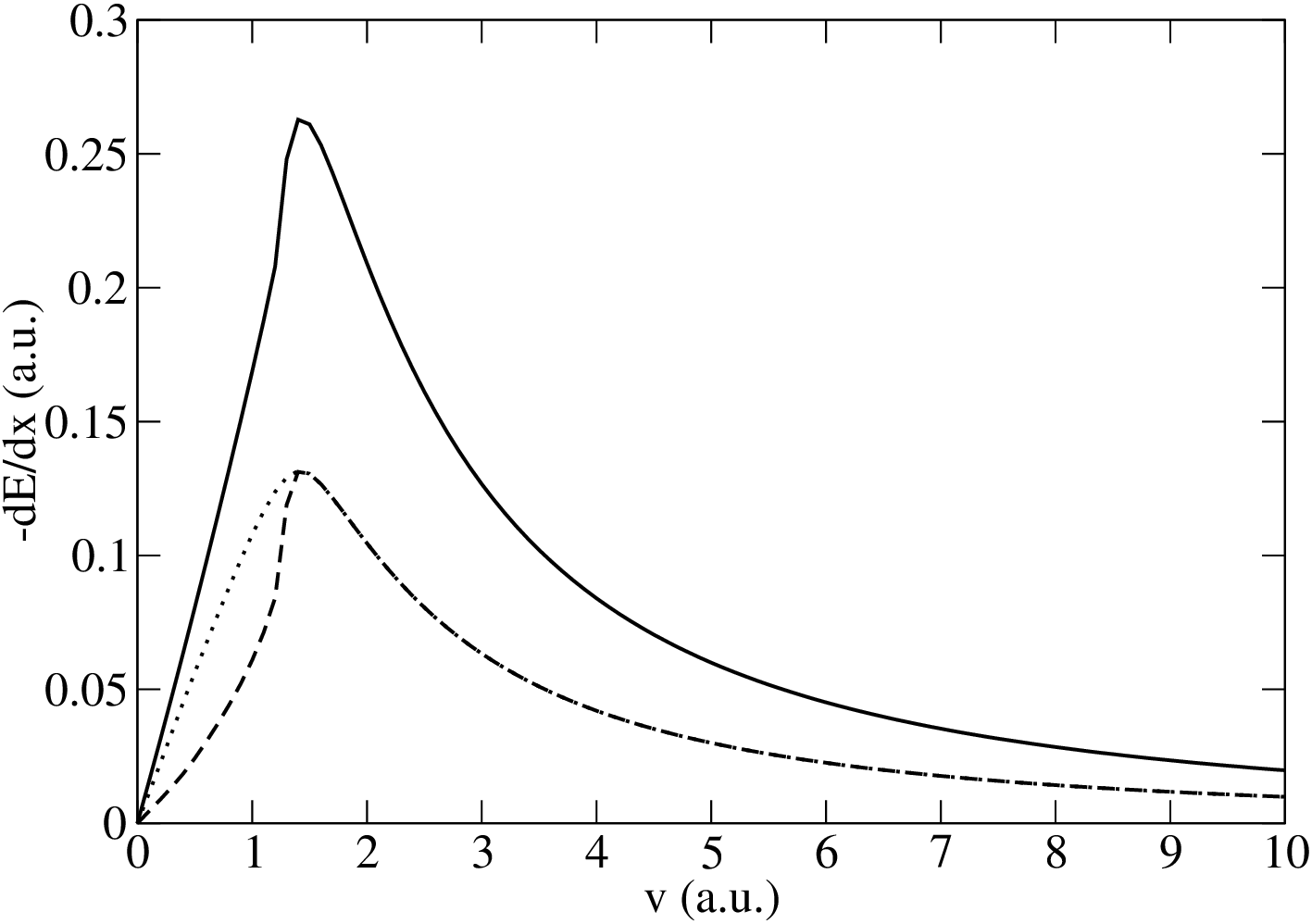}
\caption{\label{fig4}
As in Fig.~\ref{fig3}, but with the actual critical momentum $q_c$ replaced by
its low-density limit: $q_c\to\sqrt{2\omega_p}$.}
\end{center}
\end{figure}

For a momentum transfer that is smaller than $q_c$, both
plasmon and e-h pair excitations contribute to the full RPA energy loss, though
contributions from losses due to the excitation of e-h pairs are very small.
For $q>q_c$, however, only e-h pair excitations contribute. This is
illustrated in Fig.~\ref{fig3}, where the total $Z_1^2$ stopping power is
separated according to whether losses correspond to momentum transfers
below ($q<q_c$) or above ($q>q_c$) the critical momentum $q_c$. In
Fig.~\ref{fig4}, the total $Z_1^2$ stopping power is separated according to
whether losses come from momentum transfers below ($q<\sqrt{2\omega_p}$) or
above ($q>\sqrt{2\omega_p}$) the critical momentum $\sqrt{2\omega_p}$, which
is the low-density limit of $q_c$. In this case, there is exact equipartition
for all velocities above the stopping maximum. This equipartition is also found
to be
exact in the high-velocity limit, by using Coulomb scattering of independent
electrons with $q_{\rm min}=\omega_p/v$ or by assuming that independent
electrons are scatterd by a velocity-dependent Yukawa potential with screening
length proportional to $\omega_p/v$.

\begin{figure}
\begin{center}
\includegraphics[width=0.65\textwidth]{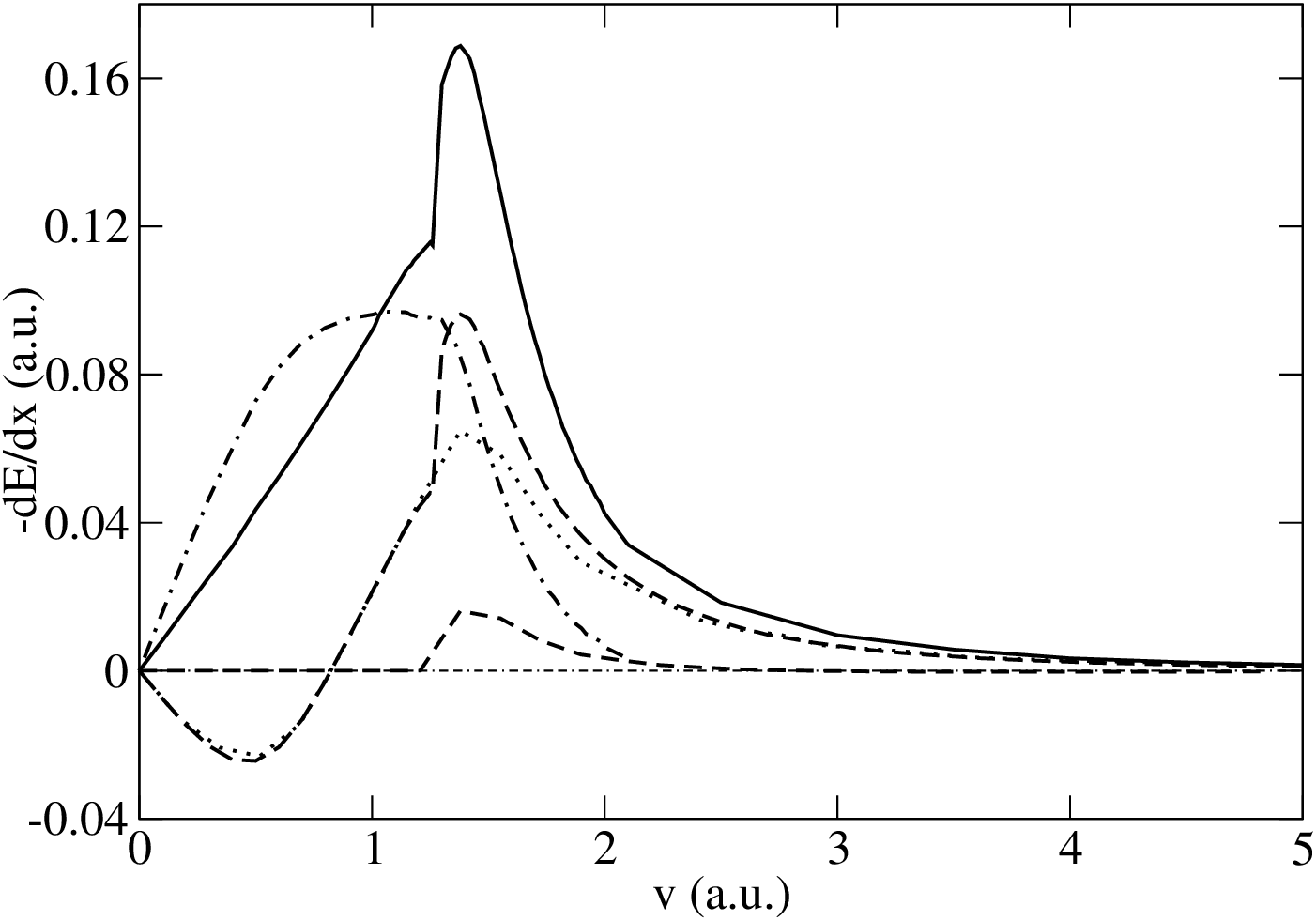}
\caption{\label{fig5}
The solid line represents the RPA $Z_1^3$ (quadratic) contribution to the
stopping power of Eq.~(\ref{eloss}), for $Z_1=1$ and $r_s=2.07$, as a
function of the velocity of the probe particle. Long-dashed and dashed-dotted
lines represent the RPA $\left[-dE/dx\right]^1$ and $\left[-dE/dx\right]^2$
contributions to the $Z_1^3$ stopping power, respectively. Short-dashed and
dotted lines represent contributions to $\left[-dE/dx\right]^1$ from the
excitation of plasmons and e-h pairs, respectively.}
\end{center}
\end{figure}

While the RPA $\left[-dE/dx\right]^2$ contribution to the $Z_1^3$ stopping
power [see Eq.~(\ref{eloss2})] is entirely due to the excitation of e-h pairs,
the $\left[-dE/dx\right]^1$ contribution to the $Z_1^3$ stopping power is
originated in the excitation of both e-h pairs and plasmons. Hence, we have
split $\left[-dE/dx\right]^1$ into these contributions and have found the
result shown in Fig.~\ref{fig5} by dotted and short-dashed lines,
respectively. This figure shows that contributions to the $Z_1^3$
effect coming from losses to plasmons is relatively smaller than in the case
of the $Z_1^2$ term (see Fig.~\ref{fig2}), especially at high velocities.
Hence, collective excitations appear to be reasonably well described with the
use of linearly screened projectile potentials. The equipartition rule, valid
within first-order perturbation (linear-response) theory cannot be extended to
higher orders in the external perturbation.

In order to account approximately for the $Z_1^3$ effect coming from both the
conduction band and the inner shells, a local-plasma approximation was used in
Ref.~\cite{pitarke1}, by assuming that a local Fermi energy can be attributed
to each element of the solid. The experimental differences between the
stopping power of silicon for high-velocity protons and
antiprotons~\cite{andersen1} were succesfully explained in this way.
Proton and antiproton stopping powers of a variety of solid targets have been
reported recently for projectile velocities near and below the stopping
maximum~\cite{andersen2}. A comparison of our theory with these experiments
requires the inclusion of charge-exchange processes, xc effects, and
higher-order nonlinear terms. Work in this direction is now in
progress~\cite{nazarov2}.

\subsubsection{Double-plasmon excitation}

\begin{figure}
\begin{center}
\includegraphics[width=0.65\textwidth]{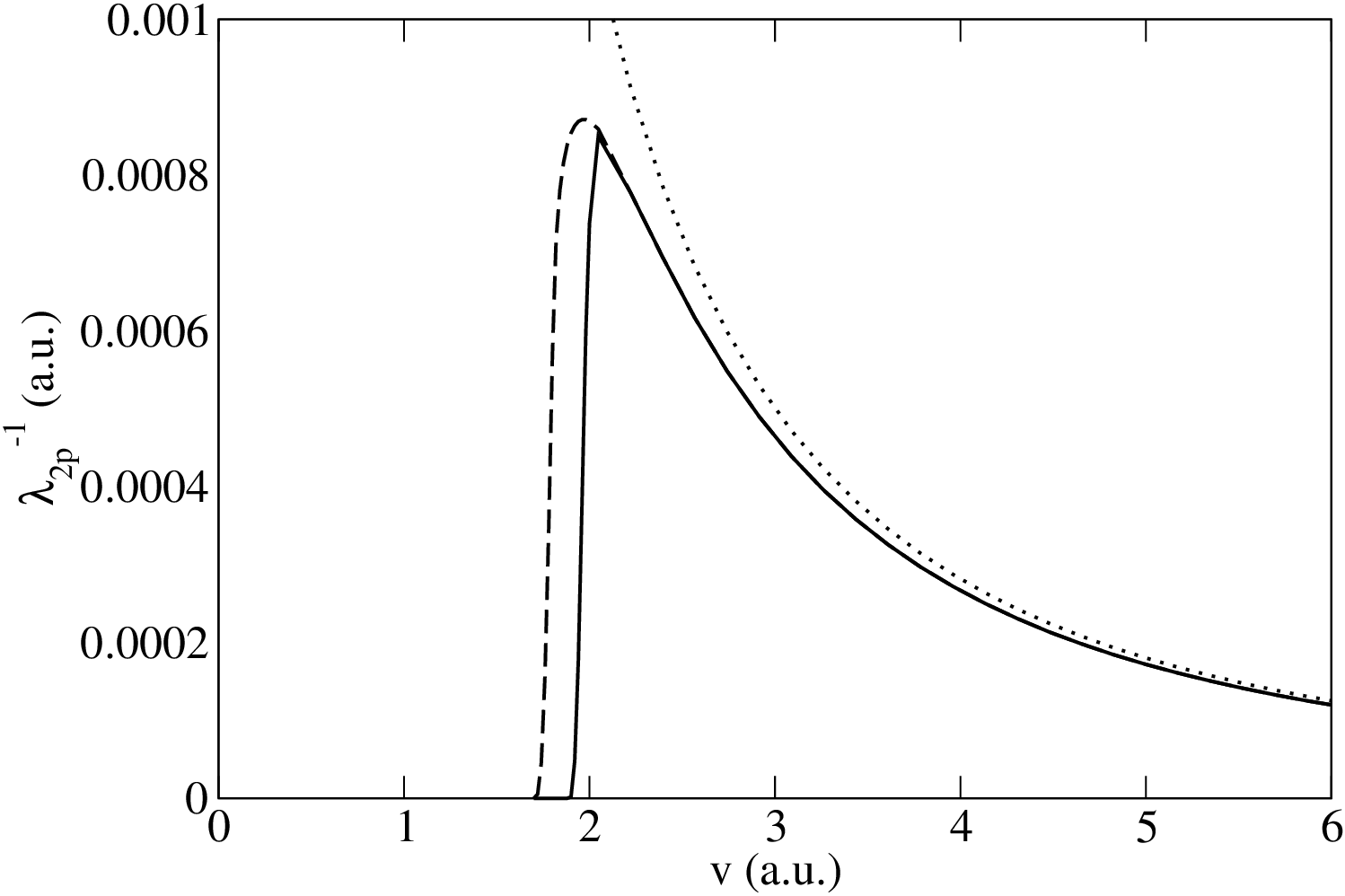}
\caption{\label{fig6}
RPA double-plasmon inverse mean free paths of electrons (solid line) and
positrons (dashed line) for $r_s=2.07$, versus the velocity of the
projectile, as obtained from Eqs.~(\ref{lambda}) and (\ref{p2p}) by either
including (electrons) or excluding (positrons) the step function
$\Theta(\omega_{{\bf v}-{\bf q}}-\varepsilon_F)$. The dotted line is the
high-velocity limit dictated by Eq.~(\ref{limit2}).}
\end{center}
\end{figure}

Single-plasmon contributions to the electron inelastic mean free path of swift
electrons have been calculated for many years, both in the high-velocity
limit~\cite{pines0} and in the full RPA~\cite{tung}. Calculations
of the double-plasmon contribution to the electron inverse mean free path
have been reported in Refs.~\cite{ashley2,pitarke2,campillo2}. Fig.~\ref{fig6}
shows
our full RPA calculation of the double-plasmon inverse mean free path of swift
electrons (solid line) interacting with a FEG, as obtained after
introducing Eq.~(\ref{p2p}) into Eq.~(\ref{lambda}). Also shown in this figure
is the double-plasmon inverse mean free path of swift positrons (dashed line),
as obtained by simply removing the step function of Eq.~(\ref{p2p}), and
the high-velocity limit dictated by Eq.~(\ref{limit}) multiplied by a factor of
2.16
(dotted line), i.e.,
\begin{equation}\label{limit2}
\lambda_{2p}^{-1}\approx 3.13\times 10^{-3}{\sqrt{r_s}\over v^2}.
\end{equation}

At high velocities of the projectile the zero-point motion of the target can be
neglected. Hence, at these velocities the effect of the Pauli restriction,
which only applies to electron probes, is negligible, and the behaviour of the
double-plasmon inverse mean free path is independent of the particle
statistics. On the other hand, it is interesting to notice that the
high-velocity formula of Eq.~(\ref{limit2}) gives an excellent account of the
full RPA result for both probe electrons and positrons in a wide range of
projectile velocities. In particular, for Al and a probe-electron energy of
$40\,{\rm keV}$ Eq.~(\ref{limit2}) yields a ratio for the double relative to
the single plasmon inverse mean free path of $1.9\times10^{-3}$, in agreement
with experiment~\cite{double2}.

\subsection{Bounded electron gas}

First, we consider a jellium slab of thickness $a$ normal to the $z$ axis,
consisting of a fixed uniform positive background of density
\begin{equation}
n_+(z)=\cases{\bar n,&$-a\leq z\leq 0$\cr\cr 0,& elsewhere,}
\end{equation}
plus a neutralizing cloud of interacting electrons of density $n(z)$.
The positive-background charge density $\bar n$ is expressed in terms of
the Wigner radius $r_s$ [$1/\bar n=(4\pi/3)(r_s\,a_0)^3$], which we take to
be $r_s=2.07$.

To compute the interacting RPA density-response function of Eq.~(\ref{chiz}),
we follow the method described in Ref.~\cite{eguiluz}. We first assume that
$n(z)$ vanishes at a distance $z_0$ from either jellium edge~\cite{note5}, and
expand the wave functions $\phi_i(z)$ in a Fourier sine series. We then
introduce a double-cosine Fourier representation for the density-response
function, and find explicit expressions for the stopping power of
Eq.~(\ref{stop}) in terms of the Fourier coefficients of the density-response
function~\cite{aran}. We take the wave functions $\phi_i(z)$ to be the
eigenfunctions of a one-dimensional LDA hamiltonian with use of the
Perdew-Zunger parametrization~\cite{pz} of the Quantum Monte Carlo xc energy
of a uniform FEG~\cite{ca}

Finally, the stopping power of a semi-infinite FEG is obtained with the use of
the following relation:
\begin{equation}
-{dE\over dx}=
{\left[-dE/dx\right](a_n^-)+\left[-dE/dx\right](a_n)+\left[-dE/dx\right](a_n^+)\over
3},
\end{equation}
where $a_n$ is the threshold width for which the $n$th subband for the $z$
motion is first occupied, $a_n^-=a_n-\lambda_F/4$, and
$a_n^+=a_n+\lambda_F/4$,  $\lambda_F=2\pi/(3\pi^2\bar n)^{1/3}$ being the
Fermi wavelength~\cite{pe}. Converged results have been found with the use of
slabs with $n=12$, for which $a\approx 5-6\lambda_F$.

\begin{figure}
\begin{center}
\includegraphics[width=0.65\textwidth]{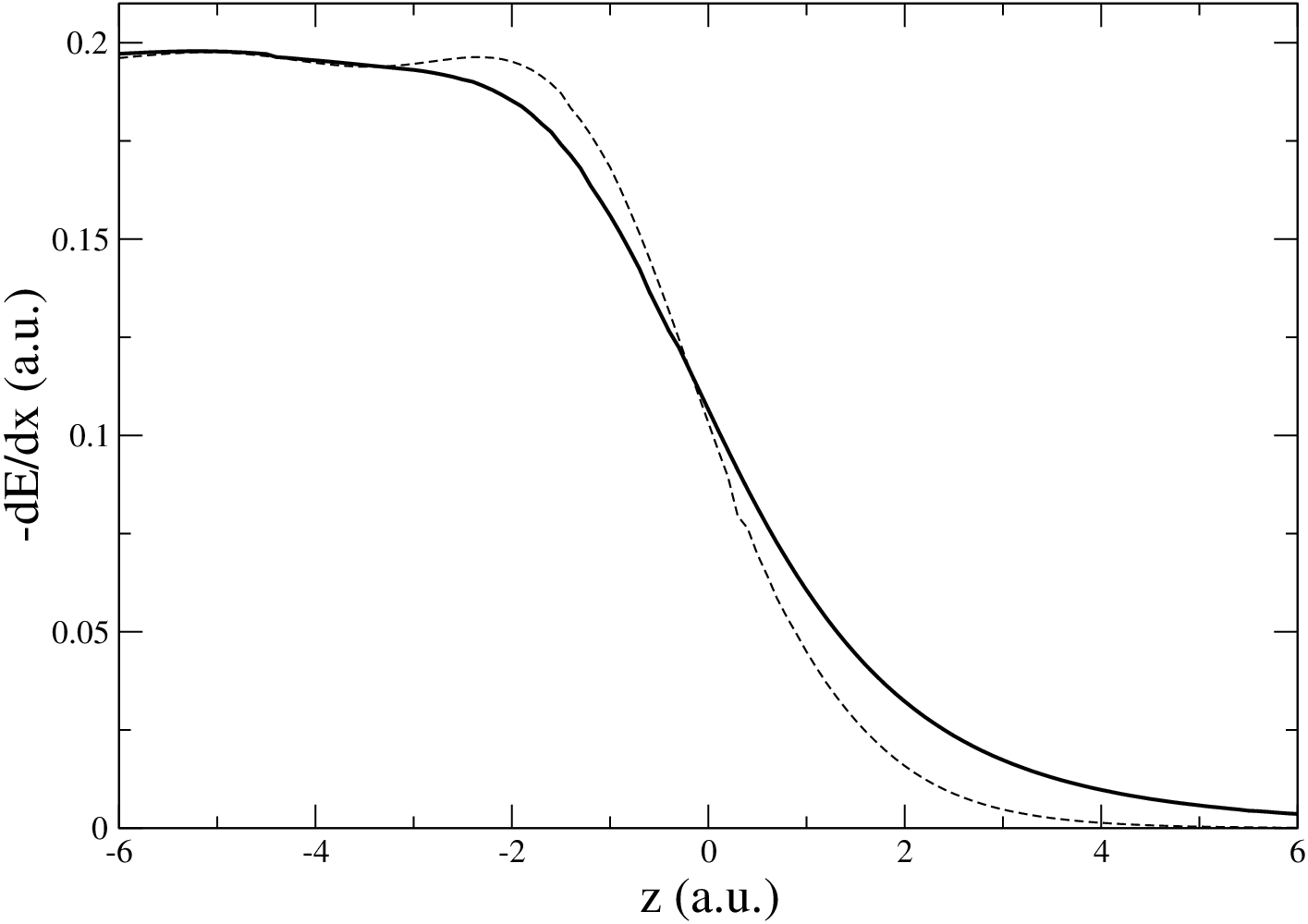}
\caption{\label{fig7}
The solid line represents the RPA stopping power of Eq.~(\ref{stop}) for
$Z_1=1$, $r_s=2.07$, and $v=2\,v_0$, as a function of the distance $z$
between the surface and the projectile. The solid is in the region $z<0$. The
dashed line represents the LDA stopping power, as obtained by assuming that
the actual stopping power of Eq.~(\ref{stop}) can be approximated by that of a
uniform FEG with the local density $n(z)$.}
\end{center}
\end{figure}

Fig.~\ref{fig7} depicts our full RPA calculation of the stopping power of a
semi-infinite FEG, as obtained from Eq.~(\ref{stop}) for protons ($Z_1=1$)
moving with speed $v=2\,v_0$ parallel to the surface. In the interior of the
solid, where the electron density is taken to be constant, the stopping power
coincides with the $Z_1^2$ (linear) RPA stopping power of a uniform FEG (see
Fig.~\ref{fig1}). Outside the solid, the stopping power decreases with the
distance $z$ between the surface and the probe-particle trajectory. Also
plotted in Fig.~\ref{fig7} is the result of assuming that the stopping power
for a charged particle that moves at a distance $z$ from the surface can be
approximated by that of a uniform electron gas with the local density
$n(z)$~\cite{note6}. Fig.~\ref{fig7} clearly shows that this often-used
local-density approximation yields an inaccurate description of the
position-dependent stopping power, due to the intrinsic nature of
surface-induced single and collective excitations not present within this
approach.

\begin{figure}
\begin{center}
\includegraphics[width=0.65\textwidth]{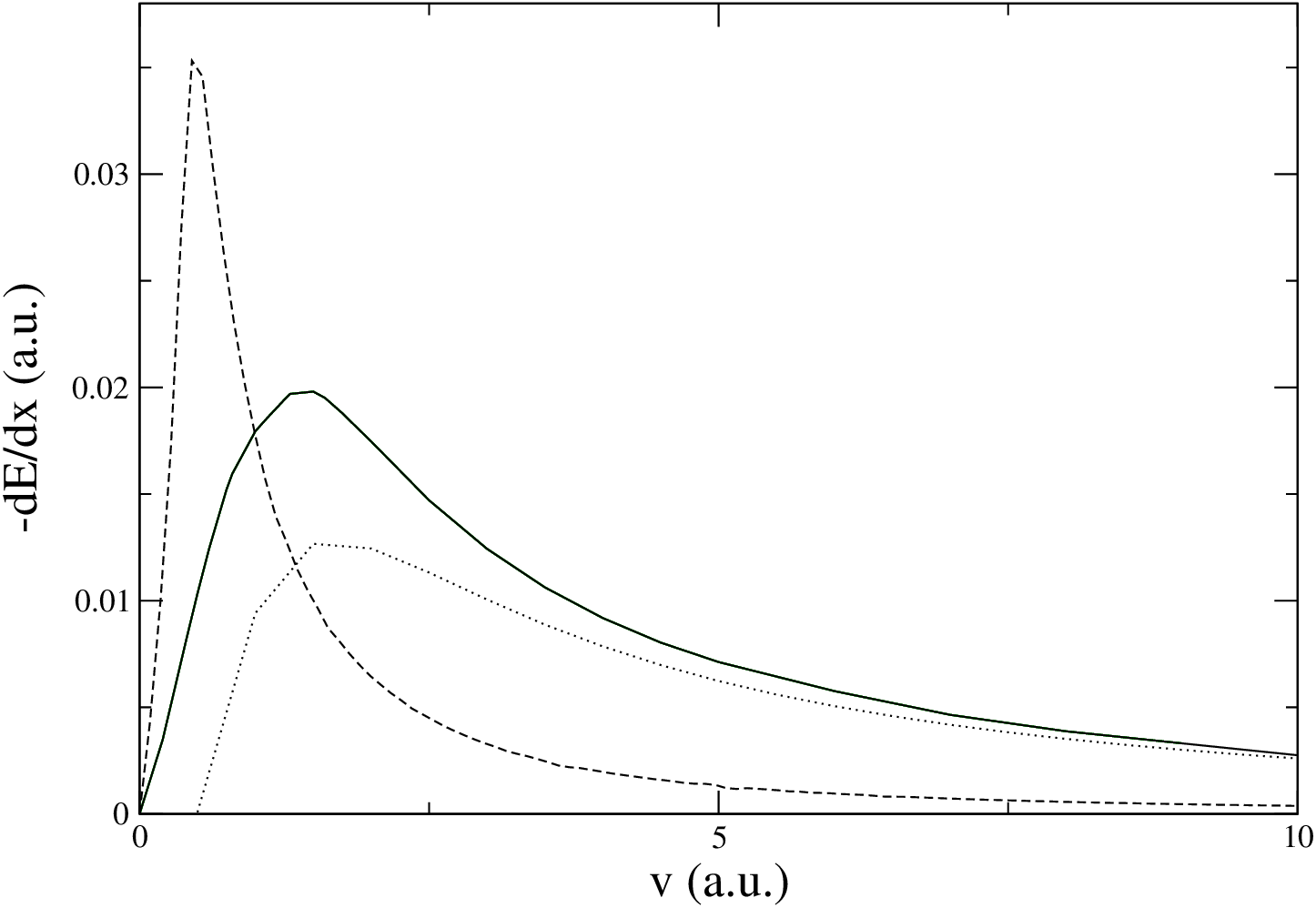}
\caption{\label{fig8}
The solid line represents the RPA stopping power of Eq.~(\ref{stop}) for
$Z_1=1$, $r_s=2.07$, and $z=3\,a_0\approx\lambda_F/2$, as a function of the
projectile velocity. The dashed line represents the LDA stopping power, as
obtained by assuming that the actual stopping power of Eq.~(\ref{stop}) can be
approximated by that of a uniform FEG with the local density $n(z)$. The dotted
line is the high-velocity limit dictated by Eq.~(\ref{stopcl}).}
\end{center}
\end{figure}

As the velocity increases, the energy-loss spectrum of charged particles
moving far from the surface into the vacuum is dominated by long-wavelength
excitations. In this limit, Eq.~(\ref{stop}) yields the classical stopping
power dictated by Eq.~(\ref{stopcl}). This is illustrated in Fig.~\ref{fig8},
where the velocity-dependent RPA stopping power of Eq.~(\ref{stop}) is
represented together with the classical result [see Eq.~(\ref{stopcl})] for a
proton ($Z_1=1$) moving in the vacuum at a constant distance
$z=3\,a_0\approx\lambda_F/2$ from the surface. At low velocities, the
energy-loss spectrum is dominated by intermediate and short wavelength
excitations, even far from the surface into the vacuum, and a combination of
the actual electronic selvage at the surface with the intrinsic nature of
surface-induced excitations increases the energy loss with respect to that
predicted by Eq.~(\ref{stopcl}). At high velocities, the energy-loss spectrum
is dominated by surface-plasmon excitations and the full RPA stopping power
nicely converges with the classical result. As in Fig.~\ref{fig7}, the
local-density approximation is also represented in this figure, showing that
this often-used approximation cannot account for the energy loss originated in
surface-induced excitations, not even at low velocities where the energy loss
is entirely due to the excitation of e-h pairs.

\subsection{Periodic crystals}

In the case of periodic crystals, we first expand the one-electron $\phi_{{\bf
k},n}({\bf r})$ eigenfunctions in a plane-wave basis,
\begin{equation}\label{eq10}
\phi_{{\bf k},n}({\bf r})={1\over\sqrt\Omega}\sum_{\bf G} u_{{\bf k},n}({\bf
G}){\rm e}^{{\rm i}({\bf k}+{\bf G})\cdot{\bf r}},
\end{equation}
with a kinetic-energy cutoff that varies from 12 Ry in the case of Al ($\sim
100\,{\bf
G}$-vectors) to 16 Ry in the case of Si ($\sim 300\,{\bf G}$-vectors). The
coefficients $u_{{\bf k},n}$ are evaluated by solving the Kohn-Sham equation
of DFT in the LDA with use of the Perdew-Zunger parametrization~\cite{pz} of
the Quantum Monte Carlo xc energy of a uniform FEG~\cite{ca}. The electron-ion
interaction is based on the use of an {\it ab initio} non-local,
norm-conserving ionic pseudopotential~\cite{hamann}. Then, Eq.~(\ref{eq9}) is
used to evaluate the Fourier coefficients of the noninteracting
density-response function, and a matrix equation [Eq.~(\ref{eq28})] is solved
for the Fourier coefficients of the interacting RPA density-response function.

Here we present new calculations for the random stopping power of Al and Si,
which represent an extension of existing first-principles
calculations~\cite{igor0,igor1} by allowing transitions with larger values of
the momentum transfer.

\subsubsection{Random stopping power}

\begin{figure}
\begin{center}
\includegraphics[width=0.65\textwidth]{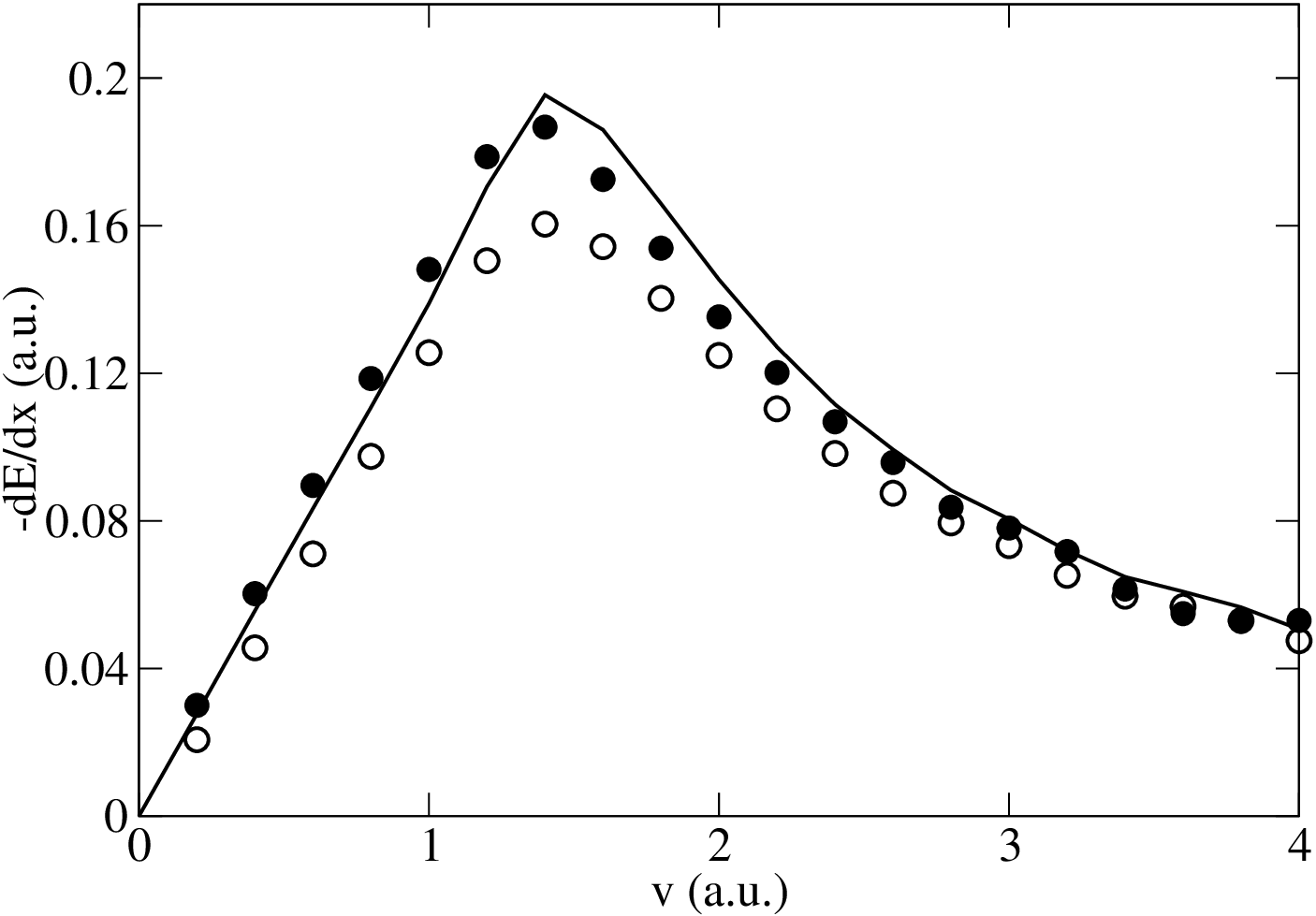}
\caption{\label{fig9}
First-principles RPA calculation of the random stopping power of valence
electrons in Al (solid circles) and Si (open circles) for protons and
antiprotons ($Z_1^2=1$), versus the projectile velocity, as obtained from
Eq.~(\ref{random}). These results have been found to be rather insensitive to
the
choice of the direction of the projectile velocity. The solid line represents
the $Z_1^2$ (linear) stopping power of a uniform FEG with $r_s=2$.}
\end{center}
\end{figure}

Fig.~\ref{fig9} shows our first-principles first-order (linear-response) RPA
calculation of the random stopping power of valence electrons in Al (solid
circles) and Si (open circles) for protons and antiprotons ($Z_1^2=1$), as
obtained from Eq.~(\ref{random})~\cite{note7}. Since the electron density of
valence electrons in Al ($r_s=2.07$) and Si ($r_s=2.01$) is nearly the same,
within a FEG model of the solid the stopping powers of Al and Si
are expected to coincide (solid line of Fig.~\ref{fig9}). Nevertheless, our
first-principles calculations indicate that this is not the case. At low
projectile velocities (where only valence electrons contribute to the energy
loss of the projectile), the stopping power of Si is considerably smaller than
that of Al, in agreement with available measurements of the stopping power of
this materials for either protons~\cite{ziegler,bauer} and
antiprotons~\cite{andersen2}. While the stopping power of Al is found to be
slightly larger than in the case of a FEG, the band gap of Si yields a
stopping power of this material that is smaller than in the case of a FEG.

At high velocities, well above the stopping maximum, the sum over the frequency
$\omega$ in Eq.~(\ref{random}) can be replaced by an integration over all
positive frequencies, and the sum rule
\begin{equation}\label{sum}
\int_0^{\infty}{\rm d}\,\omega\,\omega{\rm Im}K_{{\bf G},{\bf G}'}^{-1}({\bf
q},\omega)=2\pi^2\,n_{{\bf G}-{\bf G}'}
\end{equation}
($n_{{\bf G}}$ represents the Fourier components of the density, $n_{{\bf
G}=0}$ being
the average electron density $n$ of the crystal) yields a stopping power which
depends on $n$ but not on the details of the band-structure of the target
material:
\begin{equation}
\left[-{{\rm d}E\over {\rm d}x}\right]_{\rm
random}\sim{4\pi Z_1^2\over v^2} \,n\,\ln{2v^2\over\omega_p}.
\end{equation}
Hence, at high velocities the stopping powers of valence electrons in Al and
Si both coincide with that of a FEG with the same electron density (see
Fig.~\ref{fig9}).

While at low velocities the contribution to the total energy loss due to
excitation of inner-shell electrons is negligible, at velocities larger than
the Fermi velocity it is necessary to allow for this contribution. The cross
sections for the ionization of inner shells in Al were obtained by Ashley {\it
et al.}~\cite{ashley} in the first-Born approximation utilizing atomic
generalized oscillator strength functions. By adding the contribution from
core electrons to that of valence electrons (this contribution was calculated
within a FEG model of the solid) these authors found a nice agreement with
experiment. Good agreement with experiment was also shown in
Ref.~\cite{komarov} by adding to the valence-electron energy loss of a FEG with
$r_s=2.01$ the energy loss from core electrons in Si as taken from Walske's
calculations~\cite{walske}.

\section{Summary and conclusions}
\label{sec:summary}

We have presented a survey of current investigations of various aspects of the
interaction of charged particles with solids.

In the framework of many-body perturbation theory, we have studied the
nonlinear interaction of charged particles with a free gas of interacting
electrons. We have presented general procedures to calculate the nonlinear
potential induced by charged particles moving in an inhomogeneous electron
system, the $Z_1^3$ contribution to the stopping power of a FEG, and
double-plasmon excitation probabilities.

Our calculations for the RPA stopping power of a FEG indicate that for
velocities smaller than the Fermi velocity the stopping power is, up to third
order in the projectile charge, a linear function of the velocity of the
projectile. Our calculations also indicate that the high-velocity limit
dictated by Eq.~(\ref{limit0}) gives an excellent account of the full RPA
result in a wide range of projectile velocities. By assuming that a local
Fermi energy can be attributed to each element of the solid target, the
experimental differences between the stopping power of silicon for
high-velocity protons and antiprotons~\cite{andersen1} were succesfully
explained in Ref.~\cite{pitarke1}.

New calculations of the various contributions to the second-order stopping
power of a uniform FEG coming from the excitation of e-h pairs and plasmons
have been reported. We have found that the equipartition rule, valid within
first-order perturbation (linear-response) theory, cannot be extended to
higher orders in the external perturbation. We have also found that
contributions from collective excitations to the $Z_1^3$ term are small.

New RPA double-plasmon inverse mean free paths of electrons and positrons have
been evaluated, with explicit introduction of recoil and probe-particle
statistics. The high-velocity limit of Eq.~(\ref{limit2}) is found to give an
excellent account of the full RPA double-plasmon inverse mean free paths in a
wide range of projectile velocities. This formula [Eq.~(\ref{limit2})] yields
for Al and a probe-electron energy of $40\,{\rm keV}$ a ratio for the double
relative to the single plasmon inverse mean free path of $1.9\times 10^{-3}$,
in agreement with experiment~\cite{double2}.

We have reviewed existing self-consistent calculations of the energy loss of
charged recoiless particles moving parallel to a plane-bounded FEG, in the
framework of linear response theory. In the high-velocity limit and for
charged particles moving far from the surface into the vacuum the actual
stopping power is found to converge with the classical
limit dictated by Eq.~(\ref{stopcl}). However, at low and intermediate
velocities substantial changes in the stopping power are observed as a
realistic description of the surface response is considered, which leads to
the conclusion that a self-consistent description of the surface response is
necessary if one is to look at the energy loss of charged particles moving
outside a solid surface. Accurate measurements of the energy loss of protons
being reflected from a variety of solid surfaces at grazing incidence have
been reported~\cite{kimura1,kimura2,winter}. A theoretical description of
these experiments requires that the ion trajectory $z(t)$ be calculated and
energy losses from the excitation of inner shells be taken into account. Work
in this direction is now in progress.

Finally, we have extended existing first-principles calculations of the random
stopping power of valence electrons in Al and Si by allowing transitions to
larger values of the momentum transfer. We have found that at low velocities
(where losses from the excitation of inner shells is negligible) the
random stopping power of Si is considerably smaller than that of Al, though
both Al and Si have nearly the same valence-electron density. At high
velocities, band-structure effects become negligible and the random stopping
power of valence electrons in Al and Si nearly coincide.

A quantitative comparison of our theory with existing measurements of the
energy loss of antiprotons~\cite{andersen2} (which unlike protons carry no
bound states) in a variety of target materials can be achieved by combining
our first-principles calculations of the $Z_1^2$ (linear-response) stopping
power with $Z_1^3$ corrections in a FEG. Nevertheless, a comparison with
experiment still requires the inclusion of losses from the inner shells, xc
effects, and higher-order nonlinear terms. Work in this direction is now in
progress.

\section*{Acknowledgements}
\label{sec:ack}
We acknowledge partial support by the University of the Basque
Country, the Basque Unibertsitate eta Ikerketa Saila, and the Spanish
Ministerio de Ciencia y Tecnolog\'\i a.


\begin{thebibliography}{99}
\bibitem{note1} This approximation is valid for heavy charged particles, e.g.,
ions, and also for swift electrons moving with velocities that are large
compared
to those of the electrons in the medium. In the case of electrons, $Z_1=-1$.
\bibitem{note2} In usual practice, all single-particle wave functions and
energies are typically obtained by solving the single-particle Kohn-Sham
equation of density-functional theory in the so-called local-density
approximation (LDA) [see, e.g., Ref.~\cite{gross}].
\bibitem{note5} $z_0$ is chosen sufficiently large for the physical results to
be insensitive to the precise value employed.
\bibitem{note6} The use of this local-density approximation, in which one
replaces $-(dE/dx)(z)$ of Eq.~(\ref{stop}) by the stopping power of a uniform
electron gas of density $n(z)$, should not be confused with the use of an LDA
xc potential in DFT.
\bibitem{note7} Random stopping powers of valence electrons in Al
and Si have been calculated with the integrations over the momentum transfer
being extended from zero to $2.9\,q_F$ and $2.6\,q_F$ for Al and Si,
respectively.
\end{thebibliography}
\end{document}